\documentclass{aa}  

\usepackage{placeins}
\usepackage{graphicx}
\usepackage{lscape}                   
\usepackage{longtable}              
\usepackage{multirow}               
\usepackage{multicol}                
\usepackage{footnote}               
\usepackage{rotating}
\usepackage{threeparttable}
\usepackage{tabularx}
\usepackage{threeparttablex}
\usepackage{footmisc}
\usepackage{siunitx}
 
 \usepackage[varg]{txfonts}
\usepackage{latexsym,amsmath,amssymb}
\newcommand{\ha}{H$\alpha$} 
\newcommand{\msun}{M$_{\odot}$} 
\newcommand{\rsun}{R$_{\odot}$} 
\newcommand{\kms}{\,km\,s$^{-1}$} 

\begin{document}

   \title{
    A \ha~metric for identifying dormant black holes in X-ray transients}

   \author{J. Casares \inst{1,2}, M.A.P. Torres \inst{1,2} and S. Navarro Umpi\'errez \inst{1,2}
          }

   \institute{Instituto de Astrof\'\i{}sica de Canarias, E-38205 La Laguna, Tenerife, Spain
  \and 
   Departamento de Astrof\'isica, Universidad de La Laguna, E-38206 La Laguna, Tenerife, Spain\\
               \email{jorge.casares@iac.es}
             }
 \authorrunning{Casares, Torres \& Navarro Umpi\'errez}
\titlerunning {A \ha~metric for identifying BHs in XRTs}
   \date{Received 27 January 2025; accepted 3 June 2025}

  \abstract{
  Dormant black holes in X-ray transients can be identified by the presence of broad \ha~emission lines 
   from  quiescent accretion discs. Unfortunately,  short-period cataclysmic variables can also produce broad \ha~lines, 
  especially when viewed at high inclinations, and are thus a major source of contamination.
  Here we compare the full width at half maximum (FWHM) and equivalent width (EW) of the \ha~line in a sample of 
  20 quiescent black hole transients and 354 cataclysmic variables (305 from SDSS I to IV) with secure orbital periods ($P_{\rm orb}$) and find that: 
  (1)  FWHM and EW values decrease with $P_{\rm orb}$, and (2) for a given $P_{\rm orb}$ both parameters are 
  typically larger in black hole transients than in cataclysmic variables. 
  We also  compile spectral types for 17 low-mass companions in black hole  transients 
  from the literature and derive an  empirical $P_{\rm orb}-T_{\rm eff}$  calibration.  
 Using this, we conclude that the decrease in EW with $P_{\rm orb}$ is mostly driven by the dilution of the \ha~
 flux by the donor star continuum, which dominates the r-band spectrum for $P_{\rm orb}\gtrsim0.2$ d. 
At shorter periods, the larger contribution of the disc to the total r-band flux introduces significant scatter in the EWs 
due to the changing visibility of the disc projected area with binary inclination.
 On the other hand, the larger EWs observed in black holes can be explained by 
  their extreme mass ratios (which limit the fractional contribution of the companion to the 
  total flux) and the absence of a white dwarf component (important at $P_{\rm orb}\lesssim0.085$ d). 
  Finally, we present a tentative metric, based on \ha~FWHM and EW information, and provide optimal cuts 
  to select $\sim$80 \% of the black hole X-ray transients, while rejecting $\sim$78 \% of the 
 cataclysmic variables in our sample. Such a metric, combined with other multi-frequency diagnostics, 
 can help detect new dormant black hole X-ray transients in blind large-scale surveys such as H${\alpha}$WKs 
 and its pathfinder, Mini-H${\alpha}$WKs.
 } 
   

   \maketitle

\section{Introduction}
Over the past 50 years,  black hole X-ray transients (BH XRTs) have provided us with special  
windows through which to study accretion processes in extreme environments, supernova explosions, and 
the evolution of close binary systems. In recent years, other BH populations have also  been revealed 
by gravitational wave detectors \citep{abbott19} and Gaia astrometry (e.g. \citealt{gaia24}).  
n this context, XRTs remain a key benchmark for BH properties, as they are 
supposed to descend from a well-defined binary evolution channel at near 
solar metallicity, involving a common envelope phase 
(but see \citealt{naoz16}, also \citealt{burdge24}, for an alternative formation pathway in hierarchical triples). 

With only 20 dynamically confirmed cases and another $\approx50$ candidates \citep{corral16}, the 
current sample of BH XRTs is severely limited by small numbers. In addition, the sample is X-ray selected, 
and thus biased towards transients with short recurrence times and high outburst luminosities \citep{wu10,lin19}. 
Other complex X-ray selection biases may also be at play, such as high-inclination BHs concealed by disc obscuration 
\citep{narayan05, corral13},  a potential lack of short-period ($<$4 h), radiatively inefficient  BH transients 
(\citealt{knevitt14}, see also \citealt{arur18}), period gap BH XRTs with very low X-ray luminosities \citep{maccarone13}, 
or the possible existence of faint, persistent (i.e. non-transient) BH X-ray binaries with long orbital periods \citep{menou99}. 
In addition, both the Galactic distribution of XRTs and their BH masses are likely to be shaped by natal kicks \citep{gandhi20}
and obscuration by interstellar extinction \citep{jonker21}. Similarly,  while the existence of 
the so-called lower mass-gap (i.e. dearth of BHs with masses between $\simeq2-5$ \msun), appears robust against 
transient selection effects \citep{siegel23}, it depends critically on our ability to measure accurate masses 
in the presence of accretion disc contamination \citep{kreidberg12}. 
A significant step forward in the statistics of BH XRTs, with a more cleanly selected sample, is therefore fundamental 
to disentangling selection biases and advancing our knowledge of this reference population.   

In an attempt to increase the number of BH XRTs, we have developed a series of scaling relations between fundamental 
binary parameters and the properties of the \ha~emission line, formed in the accretion disc around the BH 
(\citealt{casares15,casares16,casares22}). In particular, a correlation between the \ha~full width at half maximum (FWHM) 
and the radial velocity semi-amplitude of the companion star ($K_2$) allows for the extraction of compact object mass functions 
in systems where the spectrum of the donor star is not detected (\citealt{casares15}; hereafter C15). 
This requires independent knowledge of the orbital period, which can be obtained from light curve variability. If the 
FWHM is also measured photometrically, this opens up the new concept of ‘photometric mass function’, whereby BHs can be 
searched for and weighed photometrically in large fields of view, i.e. much more efficiently than by classical spectroscopy (see C15).    
     
In \citet{casares18} (hereafter C18), we present a proof of concept of how to obtain \ha~FWHMs using three 
custom interference filters of different widths. We also propose a new strategy (H$\alpha$WKs, an acronym for 
'{\it \ha-Width Kilodegree survey}') optimised for the detection of dormant\footnote{
We consider dormant BHs to be quiescent 
BH XRTs that will eventually trigger an outburst episode. Therefore, we will hereafter use the terms, 'dormant' and 'quiescent'  
interchangeably.} BH XRTs in a 
blind survey of the Galactic plane. In addition to \ha~widths, the H{$\alpha$}WKs photometric system also provides 
equivalent width (EW) information. Furthermore, as the three filters are centred at 6563 \AA, the results are invariant to 
interstellar extinction and the spectral energy distribution of the objects. H${\alpha}$WKs was subsequently validated in 
a feasibility test, demonstrating that \ha~FWHM and EW values can be recovered within 10\% accuracy in a sample of 
quiescent BH XRTs down to at least r=22 \citep{casares-torres18}. Essentially,  H${\alpha}$WKs exploits the width of the 
Doppler-broadened \ha~line as a proxy for the deep gravitational fields of compact stars, enabling the efficient 
selection of dormant BHs. To some extent, the approach is similar to reverberation mapping techniques, whereby line widths 
from the broad line region are used to weigh super-massive BHs in active galactic nuclei (e.g. \citealt{peterson04}).  

A search for dormant 
BHs with H${\alpha}$WKs would in principle be free from X-ray selection bias, although the strategy 
does favour the detection of accreting binaries with short periods and high inclinations (i.e. large \ha~widths). The main 
source of contamination at large FWHMs is expected to come from cataclysmic variables (CVs i.e. interacting binaries 
with accreting white dwarfs), which are extremely abundant compared to BH XRTs.  
Monte Carlo simulations have shown that a cut-off at FWHM$\gtrsim$2200 \kms~removes $\approx$99.9 \% of 
all the CVs, while retaining $\approx$50 \% of the BHs (C18). The recovery of BHs with FWHM$<$2200 \kms~is quite 
challenging as the level of CV contamination increases dramatically.  The aim of this paper is to develop a metric based on 
\ha~FWHMs and EW information 
from a large sample of quiescent BH XRTs and CVs, to help separate these two populations.
The new diagnostic will provide an additional tool to identify dormant BH XRTs 
in single epoch spectra and special photometric \ha~surveys such as  H${\alpha}$WKs and its pathfinder, Mini-H${\alpha}$WKs.    

\section{Updated collection of \ha~FWHM and EWs in quiescent BH XRTs}
\label{sec:fwhm_ew}

Table~\ref{table:fwhm_ew} gives a list of FWHM and EW measurements of the \ha~line in 20 quiescent BH XRTs. 
These were obtained from a spectroscopic database collected over several epochs spanning 30 years. The errors 
reported include systematic uncertainties to account for orbital and secular (inter-epoch) variability. 
Appendix~\ref{ap:collection} gives full details of the database (broken down into different epochs), how FWHM and 
EW values were  measured, and the systematic uncertainties applied. The numbers for XTE J1650-500 should be 
treated with caution, as the only available spectra were taken just $\approx$9 months after the peak of 
the outburst, when the binary was still fading into quiescence (see \citealt{sanchez02},  
also Appendix~\ref{ap:collection} for a photometric verification). Similarly,  the GX 339-4 
off-state data from \citet{heida17} used here might not correspond to complete quiescence due to the frequent outburst 
activity characteristic of this system. It should be emphasised that the quiescent  FWHM and EW values for each 
system are quite stable over time, despite orbital and secular variability (Appendix~\ref{ap:collection}).

   \begin{table*}
      \caption[]{FWHM and EW of \ha~lines in quiescent BH XRTs.}
         \label{table:fwhm_ew}
\centering                         
\begin{tabular}{l l c c c c c}       
\hline\hline                
            Source  &  $P_{\rm orb}$ & spectra & epochs   &  FWHM & EW & References \\
                        &    (d)  & (\#)    &   (\#) &  (\kms) & (\AA) & for $P_{\rm orb}$ \\
 \hline
V404 Cyg                & 6.471170(2)         & 208 & 18 &   1012$\pm$94       &  24$\pm$5    &   (1) \\
BW Cir                    & 2.54451(8)           &   98 &   4  &   1061$\pm$100    &  56$\pm$11    &   (2) \\      
GX 339-4                & 1.7587(5)             &     1 &   1  &   855$\pm$52        &  76$\pm$12  &   (3) \\      
XTE J1550-564       & 1.5420333(24)     &   34 &   2  &   1453$\pm$125    &  29$\pm$5    &   (4) \\      
MAXI J1820+070    &   0.68549(1)          &  11  &   1  &  1678$\pm$85       &  75$\pm$19  &   (5) \\      
N. Oph 77               &  0.5228(44)           &   4   &   4  &  1894$\pm$155     &  57$\pm$28  &   (6)  \\      
N. Mus 91               &  0.43260249(9)     &  88  &   5  &  1797$\pm$88       &  66$\pm$9    &   (7)  \\      
MAXI J1305-704     &  0.394(4)               &   1  &    1  &   2350 $\pm$203   &  33$\pm$6    &   (8) \\      
GS 2000+25           &  0.3440915(9)        &   2   &   2  &  2192$\pm$100    &  28$\pm$7     &   (9) \\      
A 0620-00               &  0.32301415(7)      &101  &   6  & 1938$\pm$94       &  58$\pm$12   &  (10) \\      
XTE J1650-500 \tablefootmark{a}     &  0.3205(7)             &   1  &   1   &  1898$\pm$121     &   9$\pm$2     &  (11), (12) \\      
N. Vel 93                 &  0.285206(1)         &    1  &  1   &  2055$\pm$124     &  63$\pm$10   &  (13) \\      
N. Oph 93               &  0.278(8)               &    1  &  1   &  2267$\pm$194     &  89$\pm$15   &  (14) \\      
XTE J1859+226      &  0.276(3)               &  48  &  4   &  2341$\pm$145    & 108$\pm$18   &  (15) \\      
KY TrA                     &  0.26(1)                &     4  &  1   & 2167$\pm$185    &  56$\pm$18   &  (16) \\      
GRO J0422+32       & 0.2121600(2)      &    40  &  4  & 1478$\pm$55      & 289$\pm$39  &  (17) \\      
XTE J1118+480       & 0.16993404(5)     & 131  &  6  &  2732$\pm$161    &  87$\pm$17   &  (18) \\     
Swift J1753.5-0127  & 0.1358(8)            &   40  &  3   & 3502$\pm$145    & 166$\pm$17   &  (19) \\      
Swift J1357.2-0933  & 0.106969(23)      &   70  &  3   &4137$\pm$206     & 105$\pm$15   &  (20) \\      
MAXI J1659-152      & 0.10058(22)        &     1  &  1   & 3309$\pm$361    & 130$\pm$23   &  (21) \\      
                        \noalign{\smallskip}
            \hline
 \end{tabular}
\tablebib{(1)~\citet{casares19}; (2) \citet{casares09}; (3) \citet{heida17};
(4) \citet{orosz11}; (5) \citet{torres19}; (6) \citet{harlaftis97}; (7) \citet{wu15}; (8) \citet{mata21}; 
(9) \citet{harlaftis96}; (10) \citet{gonzalez17}; (11) \citet{sanchez02}; (12) \citet{orosz04}; (13) \citet{filippenko99};  
(14) \citet{casares23}; (15) \citet{yanes22}; (16) \citet{yanes24}; (17) \citet{webb00};   (18) \citet{gonzalez17}; 
 (19) \citet{yanes25}; (20) \citet{casares22}; (21) \citet{kuulkers13}.
}
\tablefoot{\\
\tablefoottext{a}{FWHM and EW values obtained when the XRT was 1.4 mag brighter than full quiescence (Appendix~\ref{ap:collection}). 
}
}
\end{table*}

\subsection{FWHM versus orbital period}

\begin{figure}
   \centering
   \includegraphics[angle=-90,width=\columnwidth]{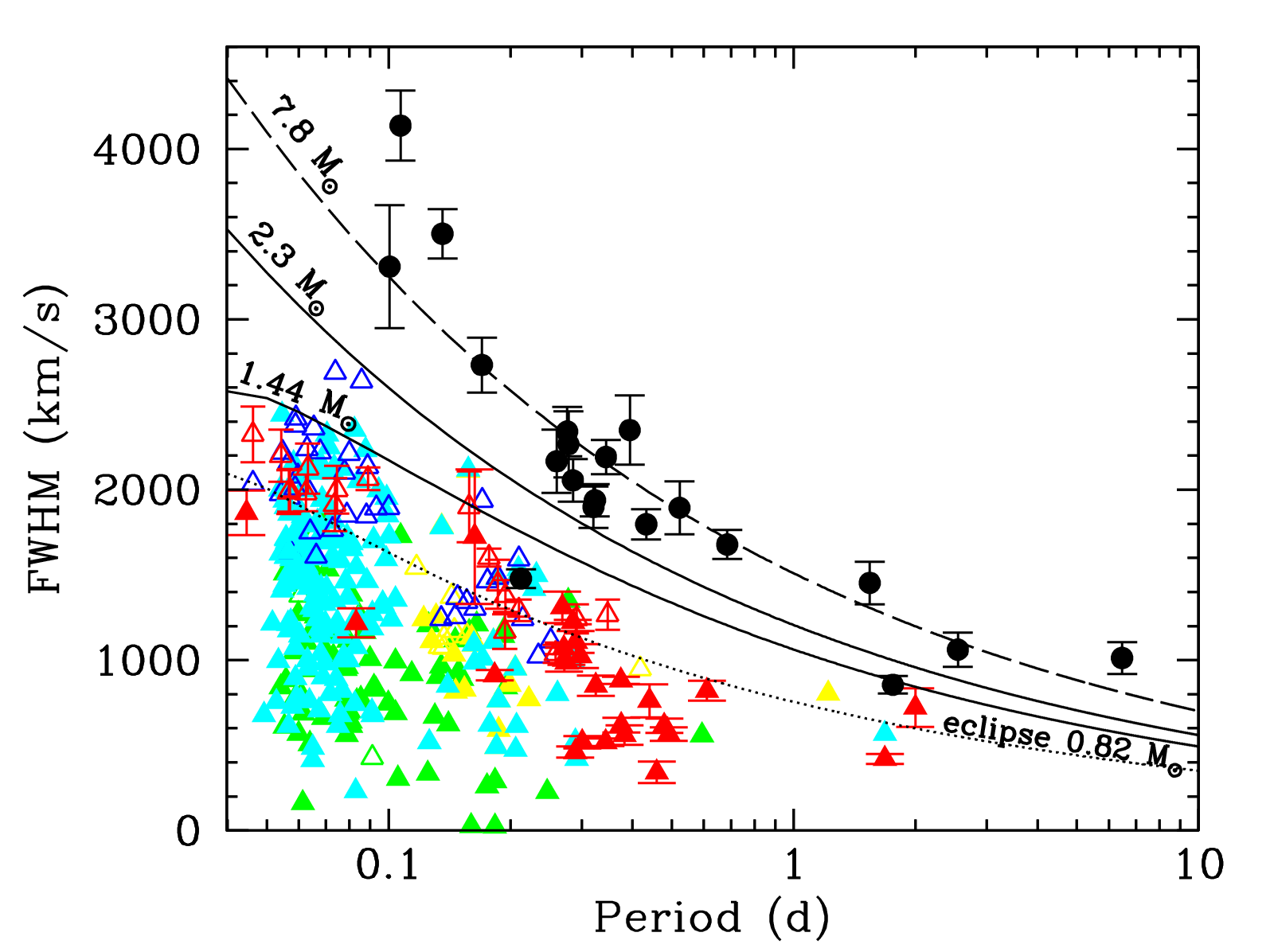}
   \caption{Distribution of BH XRTs (black circles) and CVs (triangles) in the FWHM-$P_{\rm  orb}$ plane. Open triangles indicate 
   eclipsing CVs, while filled triangles non-eclipsing CVs. Red triangles represent 43 CVs from C15 
   and blue or cyan, 
   green, and yellow triangles the 305 SDSS CVs of dwarf novae, magnetic, and nova-like types, respectively (Inight et al. 2003).
 For reference, we indicate 
  the maximum FWHM for a Chandrasekhar-mass white dwarf viewed edge-on,  and the lower limit of the FWHM for accretion disc 
  eclipses of a typical 0.82 \msun~CV white dwarf (dotted line). We also plot the maximum FWHM of a 2.3 \msun~neutron star 
  viewed edge-on. The expected track of a canonical 7.8 \msun~BH seen at 60$^{\circ}$ inclination is represented by the dashed line.
  }
              \label{fig:porb_fwhm}
    \end{figure}

Figure~\ref{fig:porb_fwhm} displays the behaviour of the FWHM as a function of the orbital period ($P_{\rm orb}$) 
for the ensemble of BH XRTs.  For comparison, we also plot FWHM values of 43 quiescent CVs (mostly dwarf novae) 
from C15, using red triangles. This sample was selected from Ritter \& Kolb's catalogue \citep{ritter03} based on 
accurate reports of the radial velocity curve of the companion star and is therefore prone to selection biases. In particular, it  
over-represents CVs with long $P_{\rm orb}$, where the companion's spectrum dominates,   
and  eclipsing short $P_{\rm orb}$  CVs, where $K_2$ is inferred through light curve modelling. 
To compensate for the uneven $P_{\rm orb}$ distribution we have added 305 CVs from the Sloan Digital Sky Survey 
(SDSS) I to IV with  secure $P_{\rm orb}$ determinations \citep{inight23}. The latter is a magnitude-limited spectroscopic 
sample and is therefore more uniformly selected. The SDSS CVs have been divided into three classes: magnetic 
(intermediate polars, polars, and pre-polars), nova-likes, and dwarf novae (including WZ Sge, SU UMa, U Gem, and Z Cam sub-types).  
Eclipsing CVs are indicated by open triangles, although the list is likely to be incomplete as some SDSS CVs lack sufficient 
photometric coverage for eclipse detection. Typical FWHM uncertainties for SDSS CVs are smaller than the symbol size 
because they are obtained from individual (single epoch) spectra. More realistic uncertainties 
(including orbital and secular variability) have been estimated at the $\approx$7 \% level in C15. 

Figure~\ref{fig:porb_fwhm} is an update of Fig. 9 from C18,  with the addition of six more BH XRTs and the 305 SDSS CVs.  
Since FWHM scales with $K_2$ (C15),  the mass function equation  $P_{\rm orb}  K_{2}^{3}/(2 \pi G) = M_{1} \sin^{3} i /(1+q)^{2}$ 
can be used to draw $P_{\rm  orb}-{\rm FWHM}$ lines for a given set of compact object mass ($M_1$), mass ratio ($q$), and 
binary inclination ($i$). As a guide, we have marked an approximate upper bound for FWHM in CVs, based on 
the Chandrasekhar mass limit and a maximum binary inclination of $i=90^{\circ}$. Here, we have also adopted the $P_{\rm orb}-q$ relation 
$q=0.73-11.55\times \exp{\left[-(P_{\rm orb}+0.39)/0.15\right]}$, derived in C18, and the $q$ dependence of the ${\rm FWHM}-K_{2}$ 
correlation (equations 5 and 6 in C15).  In addition, following C18 we have drawn a lower limit on FWHM for accretion disc eclipses of 
a representative 0.82 \msun~white dwarf \citep{zorotovic11}. An upper limit on FWHM for the case of an extreme neutron star mass, 
$M_{1}=2.3$ \msun~\citep{ruiz18, shibata19} at $i=90^{\circ}$, is also indicated. 

Figure~\ref{fig:porb_fwhm} shows that, for a given $P_{\rm orb}$, BH XRTs naturally produce broader \ha~lines than CVs due to 
the more massive central objects. As a matter of fact, BH XRTs closely follow the predicted ${\rm FWHM}-P_{\rm orb}$ curve for a 
canonical BH mass of $M_{1}=7.8$ \msun~\citep{ozel10, kreidberg12} with a typical $q=0.06$ \citep{casares16} and most probable 
inclination of $i=60^{\circ}$ (dashed line). Only GRO J0422+32 and GX339-4 fall under the 2.3 \msun~neutron star limit, placing them 
in the CV region. In the case of GRO J0422+32, this is consistent with a low-mass BH (2.7 \msun) viewed at an inclination of $i=56^{\circ}$ 
(see \citealt{casares22}). For GX 339-4, the low FWHM value could be due to either a light BH or a low inclination \citep{heida17}.
Since Fig.~\ref{fig:porb_fwhm}  is a representation of the mass function itself, it provides an efficient way to identify BHs. 
As is discussed in  C18, BHs are easily selected above FWHM$\gtrsim$2200 \kms~since CV intruders over this limit must  
have a short $P_{\rm orb}\lesssim$2.1 h (i.e. below the period gap) and are likely to be eclipsing. On the other hand, additional 
information  (such as $P_{\rm orb}$) is needed to clearly separate BHs from CVs below FWHM$\lesssim$2200 \kms.

\subsection{EW versus orbital period}

   \begin{figure}
   \centering
   \includegraphics[angle=-90,width=\columnwidth]{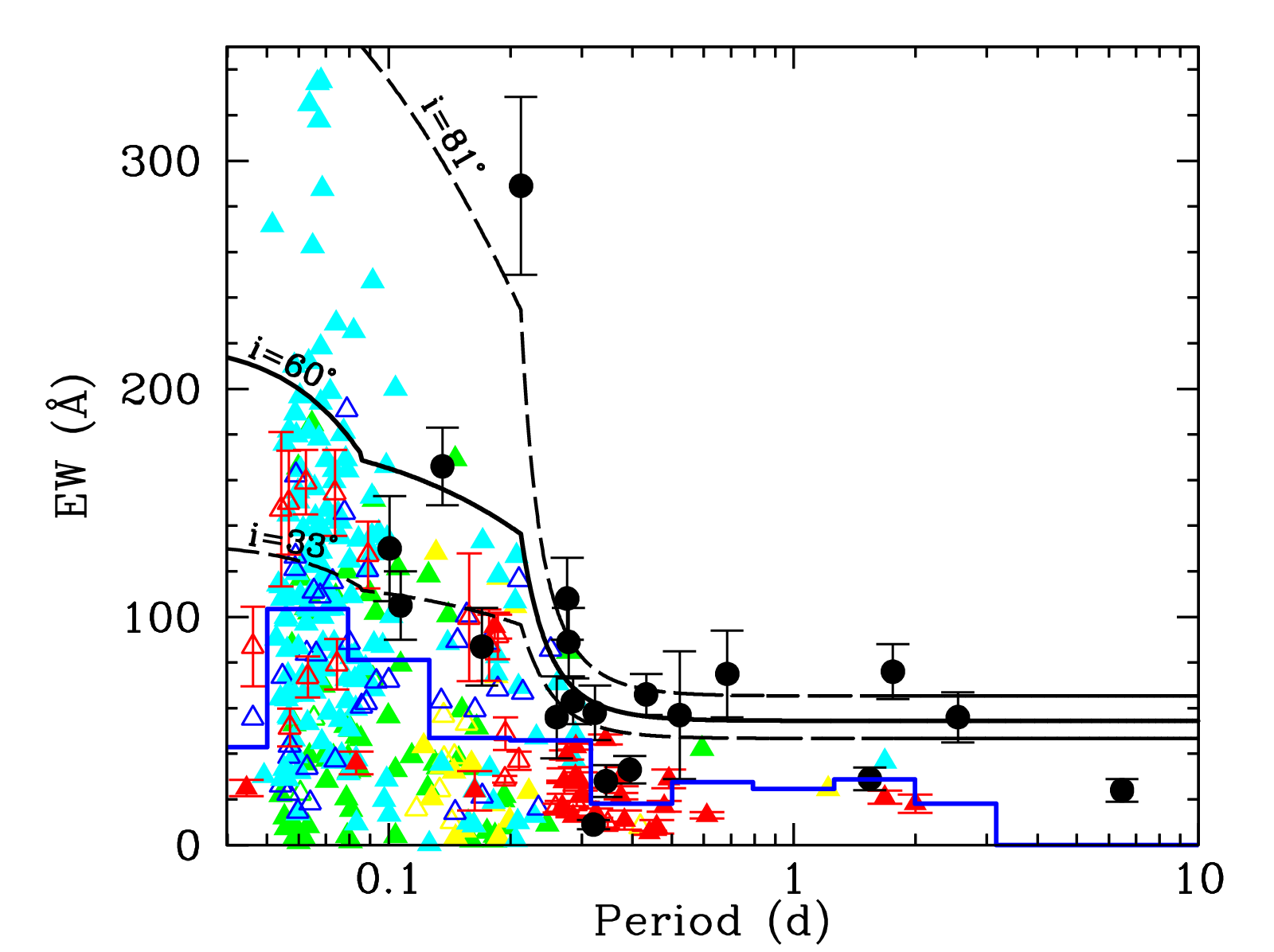}
   \caption{EW versus $P_{\rm  orb}$ for the same sample of BHs and CVs as in Fig.~\ref{fig:porb_fwhm}. For reference we plot 
   BH EW lines of constant inclination (i=33$^{\circ}$, 60$^{\circ}$ and 80$^{\circ}$), computed with our toy model simulation 
   (see Appendix~\ref{ap:model_ew} for details). The blue histogram indicates the mean EW for CVs in ten period bins.
   }
              \label{fig:porb_ew}
    \end{figure}

In Fig.~\ref{fig:porb_ew} we plot the behaviour of EWs versus $P_{\rm orb}$ for the same sample of BH XRTs and CVs 
as in  Fig.~\ref{fig:porb_fwhm}. Again, the error bars for SDSS CV EWs are not shown as they are smaller than the symbol size. 
Realistic errors, including orbital and secular variability, are estimated at the $\approx$14 \% level (see C15).  
Similar to the FWHM case, we observe that (i) EWs tend to increase with decreasing $P_{\rm  orb}$, and 
(ii) BH EWs are typically larger than CV EWs at the same $P_{\rm  orb}$. This is illustrated by the blue histogram, which 
represents the mean EW for CVs calculated in ten uniform bins of size $\Delta \log P_{\rm orb}=0.2$.  Only XTE J1650-500 
has an EW significantly below the CV mean for its $P_{\rm orb}$,  but as was mentioned earlier this is 
due to the fact that the spectra were obtained when the system had not yet reached true quiescence\footnote{
Photometry performed on the acquisition image indicates that 
the system was 1.4 mags brighter than quiescence at the time of the spectroscopic observations 
(see Appendix~\ref{ap:collection} for details). 
Assuming a constant \ha~flux, this would imply that the quoted EW is underestimated by a factor of 
$\approx$3.6.}. 
   
 To understand the EW evolution with $P_{\rm orb}$, we have built a toy model that generates synthetic EWs based on the 
 Roche geometry and the continuum blackbody radiation associated with the donor star and the accretion disc. The 
 \ha~emission is set to arise from the surface of the disc and it is assumed to be optically thin. A key ingredient of 
 the model is the $P_{\rm orb}$ dependence of the donor effective temperature, which we have calibrated in 
 Appendix~\ref{ap:teff_porb} using eq.~\ref{eq:teff_porb}. The latter was derived from a compilation of empirical spectral types 
 of BH donors (Table ~\ref{table:b1}) and CV donors from \cite{knigge06}. In the case of BH binaries we fixed $q=0.06$ 
 (\citealt{casares16}, also Table~\ref{table:a4}), while for CVs we adopted the aforementioned exponential increase with $P_{\rm  orb}$, 
 i.e. $q=0.73-11.55\times \exp{\left[-(P_{\rm orb}+0.39)/0.15\right]}$.  In the case of CVs, we also added an extra blackbody 
to account for the emission from the white dwarf. Further details of the modelling are given in Appendix~\ref{ap:model_ew}. 

 The black lines shown in Fig.~\ref{fig:porb_ew} represent the BH EWs predicted by our model for three different inclinations: 
 33$^{\circ}$, 60$^{\circ}$, and 80$^{\circ}$, i.e. the median and extremes containing 68 \% of the values for an isotropic distribution 
 of orientations. We observe that, despite the very rough approximations involved, our model is able to provide a qualitative 
 description of the behaviour of the BH EWs with  $P_{\rm  orb}$. Overall, the increase in BH EWs with decreasing $P_{\rm  orb}$ 
 is explained by the reduction in donor stellar flux caused by the drop in temperature, which becomes more pronounced at 
 $P_{\rm  orb}<$ 0.3 d (see Fig.~\ref{ap:porb_teff}). As a consequence, the accretion disc becomes the dominant source of 
 continuum light at $P_{\rm  orb}\lesssim$ 0.2 d and, due to the foreshortening caused by binary inclination, a wider range 
 of EW values is expected at these short orbital periods. 
 
As is shown in Appendix~\ref{ap:model_ew}, the model is also able to reproduce the larger EWs observed in BHs 
simply by invoking their smaller $q$ values (see Fig.~\ref{ap:porb_ew}). This has the effect of reducing the relative 
contribution of the donor star to the continuum flux, increasing the EW of the \ha~line. Only at very short periods 
below the period gap ($P_{\rm  orb}\lesssim 0.085$ d) do CV mass ratios become comparable to the ones of BHs, but 
then the white dwarf contribution starts to dominate over the donor's continuum in the \ha~region, keeping the EW values of CVs below 
the EWs of BHs.
Obviously the model is far too simplistic to explain the EW scatter seen in systems 
with similar $P_{\rm  orb}$, mass ratios and inclinations. Different accretion disc structures and the intrinsic variability 
in the continuum and/or the \ha~flux (perhaps in response to irradiation, e.g. \citealt{hynes02,hynes04}) are also likely to play
an important role in the observed EW values. 

We have so far avoided quiescent XRTs with neutron stars in this paper. This is because the 
number of systems with available quiescent FWHM and EW measurements is limited to Cen X-4 and XTE J2123-058 
(see Table 3 in \citealt{casares15}). A third obvious candidate, Aql X-1, is unfortunately contaminated by a very bright interloper  
\citep{chevalier99, mata17}, which biases the EW and FWHM determination.  
Cen X-4 and XTE J2123-058 fall in regions of the 
 $P_{\rm  orb}$-FWHM and  $P_{\rm  orb}$-EW diagrams populated by CVs, and therefore cannot be unambiguously identified. 
 This is not surprising given the comparable mass ratios and compact object masses of the two populations.

\section{A \ha~metric for selecting quiescent BH XRTs}

\begin{figure}
   \centering
   \includegraphics[angle=0,width=\columnwidth]{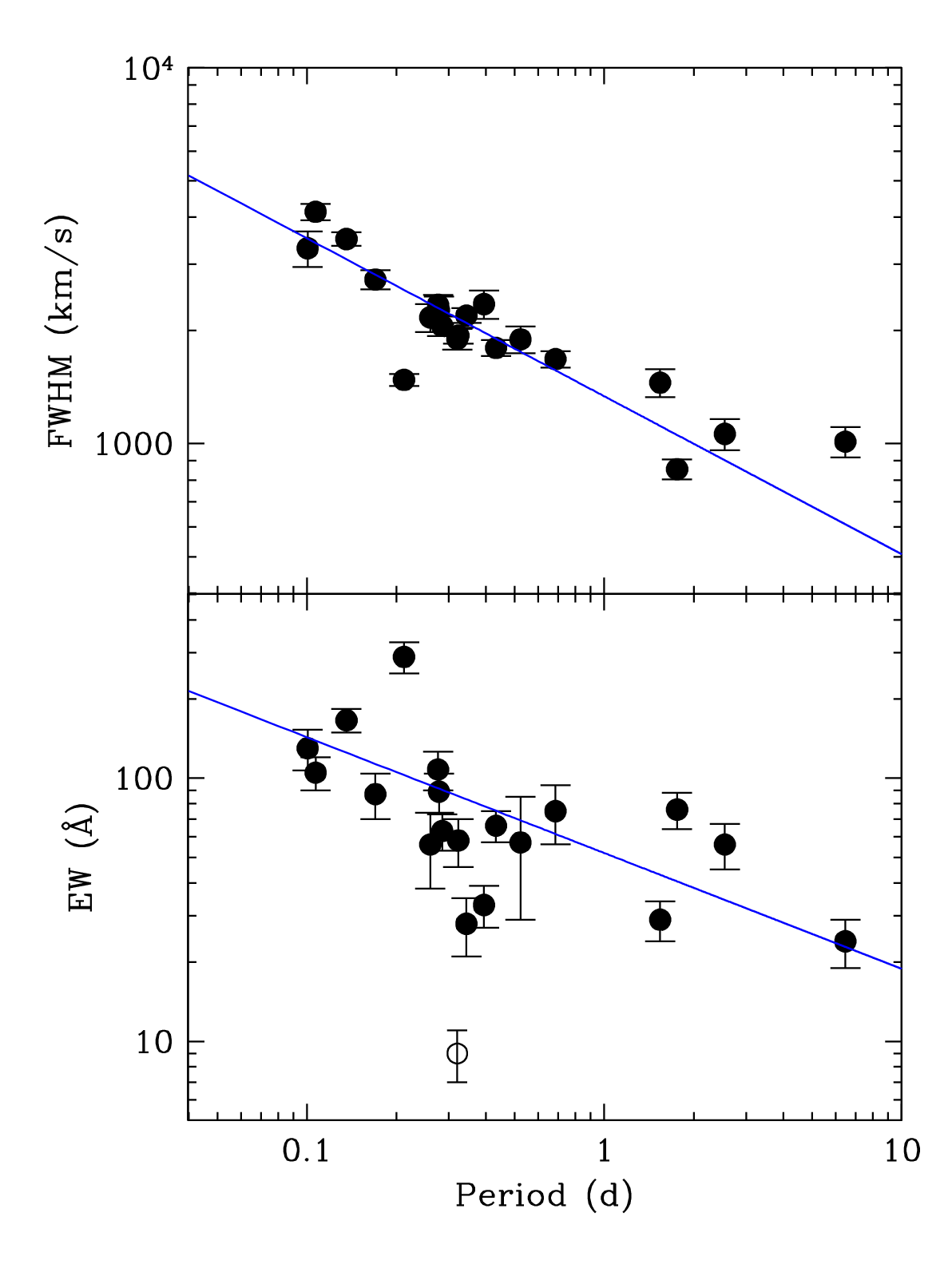}
   \caption{Power-law fits to the orbital dependence of the FWHM and EW in quiescent BH XRTs. 
   The EW of XTE J1650-400 (open circle) has been masked from the fit.}
              \label{fig:power_law_fit}
    \end{figure}

In this section we exploit the orbital period dependence of the FWHM and the EW of \ha~lines in BH XRTs to facilitate 
the selection of new dormant BHs. A least-squares power-law fit to the  BH FWHM data versus  $P_{\rm  orb}$ yields 
FWHM$=1337~P_{\rm  orb}^{-0.42}$, with FWHM expressed in units of 
kilometers per second 
and $P_{\rm  orb}$  in days.The choice of 
a power law is physically motivated by the mass  function equation, i.e. FWHM $\propto K_{2} \propto P_{\rm  orb}^{-1/3}$. 
We have 
not considered the error bars in the fit to avoid this 
being dominated by points with a low FWHM, and thus small fractional errors. 
Similarly, a power-law fit to the EWs gives  EW$=52~P_{\rm  orb}^{-0.44}$, with EW in units of 
angstroms. 
Adopting other functional forms, such as an exponential decay or polynomial, does not lead to statistically improved fits.
Here we have masked the extremely low EW outlier of XTE J1650-400 because the binary was not in true 
quiescence. Both power-law fits to the FWHM and EW BH values are shown in Fig.~\ref{fig:power_law_fit}.

\begin{figure}
   \centering
   \includegraphics[angle=-90,width=\columnwidth]{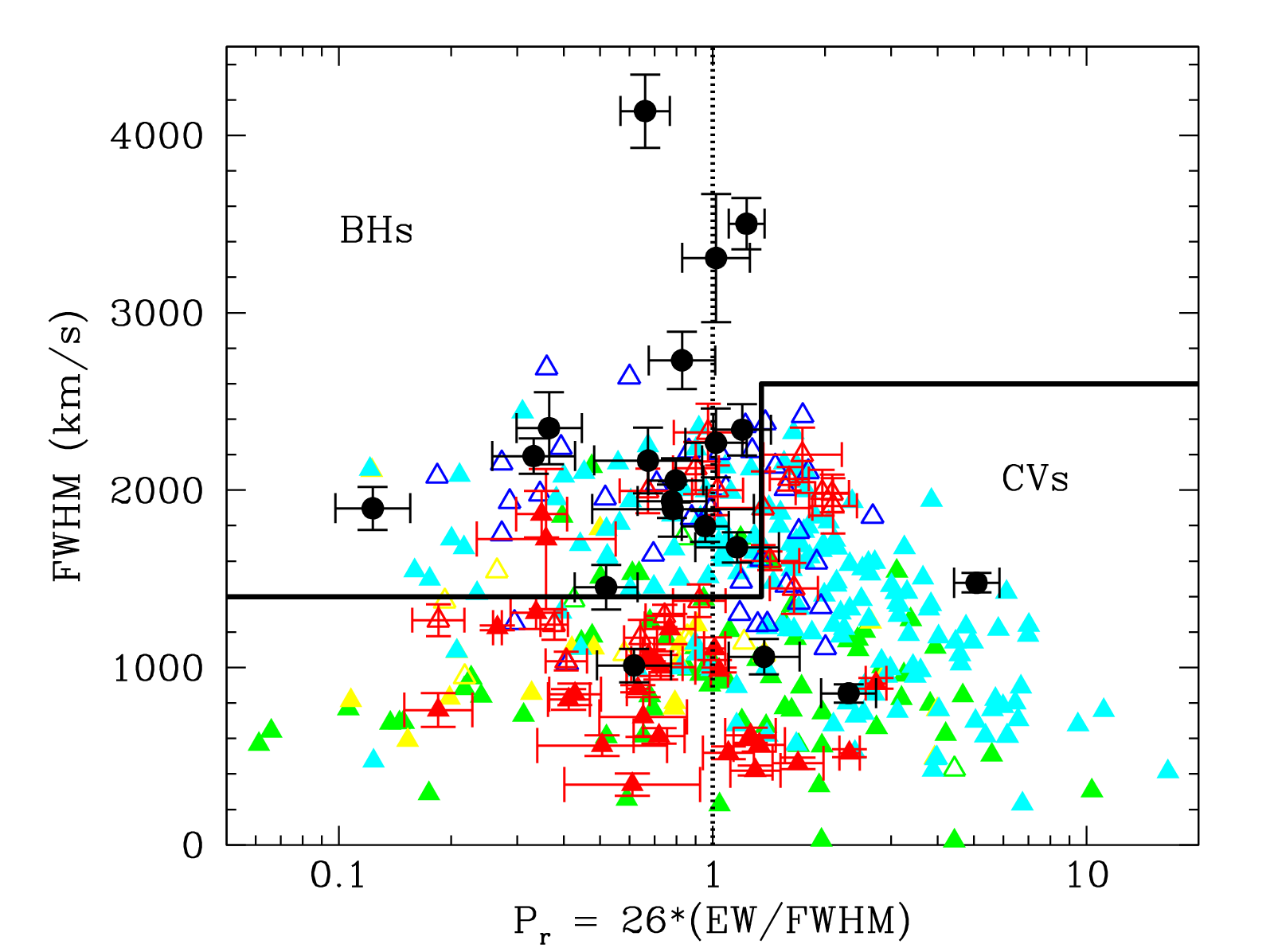}
   \caption{Diagram showing the distribution of BHs and CVs in the FWHM-$P_{\rm r}$ plane. The same symbol code 
   is used as in Figs.~\ref{fig:porb_fwhm} and \ref{fig:porb_ew}. Vertical dots mark the BH clustering line, while the thick 
   solid lines delineate the  optimal region for separating BHs from CVs.}
              \label{fig:metric_period}
    \end{figure}

The previous expressions provide rough estimates of BH orbital periods based on FWHM and EW 
information alone. We refer to these as  $P_{\rm FWHM}$ and $P_{\rm EW}$, respectively. 
Given the scatter in the power-law fits, the difference between these estimates and the true $ P_{\rm  orb}$ 
can be significant. However, since both $P_{\rm FWHM}$ and $P_{\rm EW}$ are expected to track $P_{\rm  orb}$ 
better in BHs than in CVs, we decided to choose the ratio  $P_{r} \equiv (P_{\rm FWHM}/P_{\rm EW})^{0.43}$ 
as a suitable metric for BH selection, which we now approximate as  $P_{r} \approx 26~ ({\rm EW/FWHM})$. 
Figure~\ref{fig:metric_period} depicts  $P_{r} $  against FWHM for our BHs and the ensemble of CVs. 

Since BHs are expected to cluster around $P_{r} \approx 1$, it thus seems convenient to choose vertical 
bands of different widths, centred at  $P_{r}=1$, to optimise their selection against CVs.  As is shown in 
Table~\ref{table:bh_selection}, the narrower the band the more CVs are rejected, but this comes at the 
cost of selecting fewer BHs too. Interestingly, the distribution of BHs  appears to be skewed towards lower 
$P_{r}$ values, with very few BHs also found under FWHM$=1400$ \kms. Furthermore, dynamical arguments 
based on the Chandrasekhar mass limit indicate that CVs cannot produce \ha~lines broader than 
$\simeq$2600 \kms~(see Fig~\ref{fig:porb_fwhm}, also C15). Given the above constraints, we propose an 
optimal region for BH selection defined by the limits FWHM$>1400$ \kms~for $P_{r} < 1.35$ and 
FWHM$>2600$ \kms~for $P_{r} > 1.35$. These cuts are shown in Figure~\ref{fig:metric_period} 
and allow one to select 80 \% of the current sample of BHs, while rejecting 77 \% of the CVs. It is worth noting 
that at least 33 \% of the non-rejected CVs are eclipsing\footnote{Other non-rejected SDSS CVs may also be 
eclipsing, but have not yet been confirmed due to insufficient photometric coverage.} and therefore easily 
identified by  light curve variability. It should also be mentioned that the mere detection of eclipses provides us 
with $ P_{\rm  orb}$ information and thus, when combined with FWHM values, mass functions that constrain 
the nature of the compact star.

   \begin{table}
      \caption[]{BH selection against CV rejection using the $P_{r}$ metric.}
         \label{table:bh_selection}
\centering                         
\begin{tabular}{c c c }        
\hline\hline                 
      $P_{r}$       &  Selected BHs & Rejected CVs \\
 \hline
  0.10-10.0               & 85 \%   & 57 \% \\
  0.30-3.33               & 75 \%   & 62 \% \\
  0.50-2.00               & 65 \%  &  72 \%  \\      
  0.60-1.66               & 60 \%  &  80 \% \\      
  0.74-1.35               & 50 \%  &  89 \%  \\      
                        \noalign{\smallskip}
            \hline
 \end{tabular}
\end{table}

\begin{figure}
   \centering
   \includegraphics[angle=-90,width=\columnwidth]{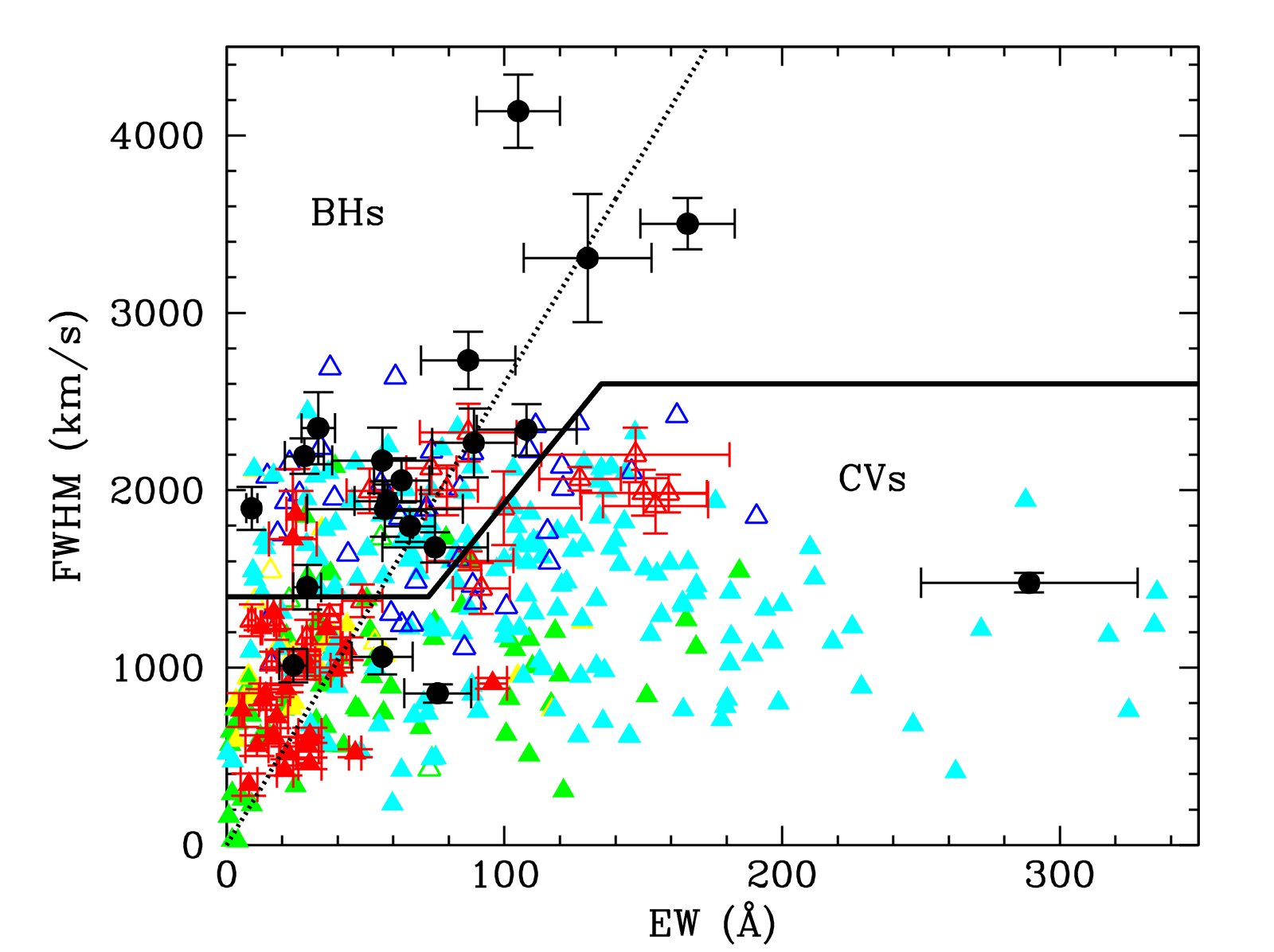}
   \caption{Distribution of BHs and CVs in the FWHM-EW plane. The BH clustering line at FWHM=26$\times$EW is 
   indicated by dots, while the solid line marks our Fig.~\ref{fig:metric_period} cuts  for optimal BH selection. 
   }
              \label{fig:metric0}
    \end{figure}

Figure~\ref{fig:metric0} displays the $P_{r}$ metric in the simpler EW-FWHM plane. Here the BHs tend 
to cluster around FWHM/EW=26 (dotted line in Fig~\ref{fig:metric0}), while the previous selection cuts are 
represented by the solid lines defined by 

\begin{equation}
\begin{array}{l l}
 {\rm FWHM}>1400 & {\rm EW} < 73  \\
 {\rm FWHM}>19.3 \times {\rm EW}  & 73 < {\rm EW} < 135  \\
 {\rm FWHM} >2600  &  {\rm EW} > 135   \\
 \end{array}
\label{eq:metric}
,\end{equation}

\noindent
with the FWHM expressed in units of 
kilometers per second and EW in angstroms. 

Obviously, the proposed cuts are based on a limited number of BHs and should be considered as preliminary. 
In any case, they can help to 
discover new 
dormant BHs under the FWHM$\approx$2200 \kms~limit proposed in C18. 
The diagnostics presented here will prove useful for current and future synoptic spectroscopic surveys (e.g. LAMOST, 
4MOST, WEAVE), in which large numbers of spectra will be collected. It will also benefit blind photometric surveys, such 
as H${\alpha}$WKs and its pathfinder (Mini-H${\alpha}$WKs), which have been tailored to extract EW and FWHM information 
from \ha-emitting objects (C18). Nevertheless, CVs have an estimated Galactic density of $\sim10^4$ kpc$^{-3}$ \citep{pala20} 
and are thus $\approx1000$ times more abundant than BH XRTs \citep{corral16}. This implies that, assuming similar absolute 
magnitudes and galactic distributions, our FWHM-EW cuts would still select $\approx$300 intruding CVs per BH. 
Clearly, additional information from other multi-frequency surveys will be crucial for unveiling BH imposters and refining 
selection diagnostics. In particular, the absence of blue/UV excess in UVEX/Galex colours (a signature of a white dwarf or 
disc boundary layer; e.g. \citealt{gansicke09}), will strengthen the possibility that new candidates are authentic dormant 
BH XRTs. Ultimately, these will need to be confirmed by dedicated follow-up spectroscopic studies.

A related question is whether
 FWHM and/or EW 
information can be used to distinguish BH XRTs that  are undergoing some level of 
accretion activity from truly quiescent systems, even if X-rays are not detected. BH XRTs are known to follow a distinct FWHM and EW pattern as they 
transition from outburst to quiescence.  
Many examples in the literature consistently show that outburst spectra evolve from being almost featureless (sometimes with weak \ha~emission embedded in 
broad absorptions) to developing increasingly stronger and broader \ha~lines. For example, during the decay of the A0620-00 discovery outburst, the EW 
increased from $\sim$5.8 to 14 \AA~and FWHM from $\sim$1100 to 1830 \kms~between 
Sept 1975 and May 1976 \citep{whelan77}. 
Other studies, focusing on the late outburst decline, suggest that the FWHM tends to approach quiescent values faster than the EW. For example,  
from November 2000 to April 2001, XTE J1118+480 promptly reached the quiescent FWHM value at $\simeq$2700 \kms, while the EWs 
showed a monotonic increase between 14 and 44 \AA, still far from the $\simeq$87 \AA~quiescent value  \citep{zurita02a, torres04}. 
In view of this, it seems tempting to conclude that \ha~EWs are a more sensitive diagnostic of disc activity than FWHMs and could perhaps be used to 
detect unusual disc activity or even precursors to new outbursts.  
Interestingly, a chance spectrum of V404 Cyg obtained 13 h before the X-ray trigger of the 2015 outburst revealed that the EW of the \ha~line was 
$\sim$15 times larger than in quiescence, while the FWHM was almost unchanged \citep{bernardini16, casares19}. 

\section{Conclusions}

   \begin{enumerate}
      \item We have compiled FWHM and EW values of \ha~emission lines in a sample of 20 quiescent BH XRTs. The BH 
      collection typically covers multiple epochs of quiescence, sometimes spanning several decades. Despite evidence of 
      orbital and secular variability, the mean FWHM and EW values are found to be rather stable (Appendix~\ref{ap:collection}).  
      The BH FWHM and EW values have been compared with the ones of a sample of 353 CVs (305 from SDSS I to IV) with known 
      orbital periods.  
     \item Our compilation shows that both FWHM and EW values decrease with $ P_{\rm  orb}$,  while, for a given $ P_{\rm  orb}$, 
     they tend to be larger in BHs than in CVs. 
      \item The larger BH FWHMs  are a natural consequence of their higher compact object masses. The larger EWs, on the 
      other hand, could be explained by a lower level of continuum flux. This stems from a combination of extreme mass ratios  
      (which limit the relative contribution of the companion to the total flux) and the absence of a white dwarf continuum. 
      \item Furthermore, we derive an empirical $P_{\rm orb}-T_{\rm eff}$  relation for companion stars in BH XRTs, which is also valid 
      for CVs (Appendix~\ref{ap:teff_porb}). From this we infer that the companion is the main source of \ha~continuum flux above 
      $\simeq$0.2 d, while the accretion disc dominates otherwise. In the case of CVs, the white dwarf also contributes to the diluting 
      continuum (especially at $P_{\rm orb}\lesssim$ 0.085 d),  further capping EW values with respect to BHs. 
       \item We finally present a tentative metric ($P_{r}$=26 EW/FWHM) for detecting dormant BHs, based on the period dependence 
       of the FWHM and EW in BH XRTs. We find that selection cuts defined by  FWHM$>1400$ \kms~for $P_{r} < 1.35$ and 
       FWHM$>2600$ \kms~for $P_{r} > 1.35$ allow us to filter out $\approx$ 77 \% of CVs, while still retaining $\approx$ 80 \% of 
       our sample of BHs. In any case, given the high Galactic density of CVs, the proposed metric needs to be combined with 
       other multi-frequency diagnostics for an efficient selection of dormant BH XRTs.         
   \end{enumerate}

\begin{acknowledgements}
We thank the anonymous referee for useful and constructive comments that helped improve the manuscript. 
JC and MAPT acknowledge support by the Spanish Ministry of Science via the Plan de Generaci\'on de Conocimiento through grants PID2022-143331NB-100 and PID2021-124879NB-I00, respectively. SNU is supported by the FPI grant PREP2022-000508, also under program PID2022-143331NB-100.  
We thank Keith Inight for sharing his database of SDSS CV spectra with us, and Cynthia Froning for sharing the SED data on A0620-00. 
We also thank Rosa Clavero, David Jones and other members of the IAC team of support astronomers for undertaking the 2016-2021 NOT 
observations of V404 Cyg during Service time. 
We dedicate this paper to the memory of Tom Marsh, creator of the molly software and a beacon in the field of compact binaries.  

\end{acknowledgements}

\begin{appendix}

\section{An extensive collection of FWHM and EW values of \ha~lines in quiescent BH XRTs}
\label{ap:collection}

We have assembled a new database of \ha~spectra of quiescent BH XRTs obtained in different epochs over 30 years. 
The present collection contains and supersedes the one reported in C15 and is presented in Table~\ref{table:a1}. 
Listed references correspond to the papers where the original spectra were first reported and/or analysed. The spectra 
were obtained with a variety of telescopes: the 10.4 m Gran Telescopio Canarias (GTC), the 10 m Keck telescope, the 8.2 m 
Very Large Telescope (VLT), the 6.5 m Magellan Clay telescope, the 4.2 m William Herschel Telescope (WHT), the 4 m 
Victor M. Blanco telescope,  the 3.9 m Anglo Australian Telescope (AAT), the 3.5 m New Technology Telescope (NTT),  
the 2.5 m Isaac Newton Telescope (INT), the 2.56 m Nordic Optical Telescope (NOT) and the 2.1 m telescope at the 
Observatorio de San Pedro M\'artir (SPM). Some data have not yet been published and will be reported elsewhere. 
These are the 2019-2021 NOT and 2022 INT epochs of V404 Cyg, the 2023 GTC epoch of MAXI J1820+070,  
the 2019 GTC epoch of XTE J1859+226 and the 
2017 GTC epoch of GRO J0422+32. 

We consider BH XRTs to be in true quiescence when there is no sign of 
X-ray activity (i.e. outbursts, mini-outbursts or reflares) and the optical flux remains 
consistent with the lowest measured values. To demonstrate that the selected spectra 
meet these criteria, we give in Table~\ref{table:a2} the dates of the quiescent periods
relevant to our data. The listed magnitudes refer either to the start of quiescence, to a 
time average or to a range of values reported over the entire period. As can be seen in Table~\ref{table:a2}, 
all the spectra correspond to quiescent epochs, including GX 339-4, which experienced a prolonged period 
of very low optical brightness in 2016 \citep{russell17}. The only exception is XTE J1650-500, as the spectroscopy of \cite{sanchez02} 
was  obtained approximately two months before the first quiescent magnitudes were reported \citep{garcia02}. To confirm this, we have 
performed PSF photometry on the acquisition images obtained by  \cite{sanchez02}, finding R=20.67$\pm$0.04, i.e. 1.4 mag brighter than 
the deepest magnitude available \citep{garcia02}. To further test  whether the latter corresponds to true quiescence, we also performed 
PSF photometry on a Sloan r-band image from the DECaPS survey\footnote{Images are publicly available at http://decaps.skymaps.info/}, 
obtained on 29 April 2017 under good seeing conditions (0.78"). We found  
r'=22.10$\pm$0.12, which is in good agreement with \cite{garcia02}. 

Following C15, we measured the FWHM of  the \ha~line by fitting a Gaussian profile plus a constant in a window of 
$\pm$10,000 \kms~centred at 6563 \AA, after masking the neighboring He I $\lambda$6678 line.  The Gaussian model 
was previously degraded to the instrumental resolution of each spectrum, and the continuum level was rectified by fitting 
a low-order polinomial.  Likewise, EW values were obtained by integrating the \ha~flux in the continuum normalised spectra.  
The entire spectral analysis was performed using routines within the MOLLY package\footnote{Molly was written by T. R. Marsh 
and is available from https://cygnus.astro.warwick.ac.uk/phsaap/software/ molly/html/INDEX.html.}. Where possible, 
FWHM and EW values were measured from single individual spectra (e.g. V404 Cyg). When the quality of individual spectra 
was too poor FWHM and EW measurements were obtained from averaged spectra (e.g. MAXI J1659-152). In some cases, 
an orbitally averaged spectrum was kindly provided by the corresponding authors (e.g. GX 339-4) while in others we had
 to digitise averaged spectra from figures in the relevant papers (e.g. the WHT 1991-1992 epoch of A0620-00). The uncertainties 
 introduced by the latter process are negligible compared to orbital and long-term variations typically observed in FWHM and EW. 

    \begin{figure}
   \centering
   \includegraphics[angle=0,width=8cm]{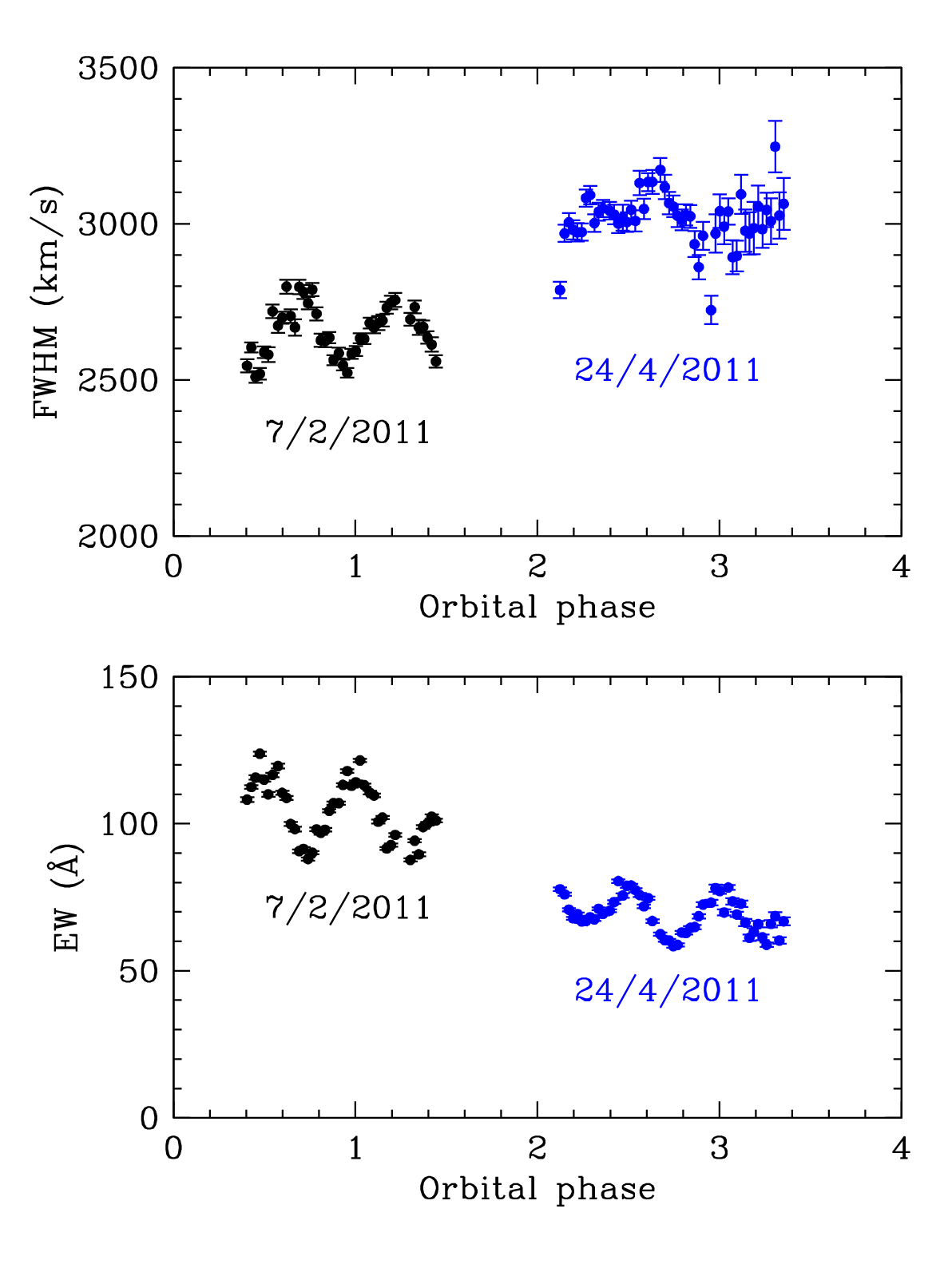}
   \caption{Orbital variation of FWHM and EW in XTE J1118+480 on the nights of February 7 and April 24 2011.}
              \label{fig:a1}
    \end{figure}
 
 Figure~\ref{fig:a1} shows examples of orbital variability in the FWHM and EW values of XTE J1118+480 at two different epochs. 
 Both parameters describe a double sine wave on each orbital cycle. The EW variations reflect changes in the optical continuum, 
 driven by the ellipsoidal modulation of the companion star (e.g. \citealt{marsh94}). On the other hand, the phasing of the FWHM 
 variations (i.e. minima at phases 0.4 and 0.9) indicate that these are likely caused by the motion of the hot-spot\footnote{In 
 semi-detached binaries with an accreting compact star a hot-spot is formed by the collision of the gas stream with the outer 
 accretion disc. For extreme mass ratios, characteristic of BH XRTs, the hot-spot crosses the observer's line of sight at orbital 
 phases $\simeq0.4$ and 0.9, with phase 0 defined as the inferior conjunction of the mass donor star.}  across the \ha~profile. 
 Consequently, when only one spectrum is available in a given epoch (e.g. the Keck 2004 spectrum of XTE J1118+480) or 
 its phase coverage is very limited (e.g. the WHT 1993 epoch of V404 Cyg), statistical uncertainties in FWHM and EW 
 measurements were increased by adding quadratically a systematic error to account for the effect of orbital variability. 
We call these 'orbital' systematic errors $\sigma({\rm FWHM})_{\rm orb}$ and  $\sigma({\rm EW})_{\rm orb}$, 
respectively. 

\begin{figure}
   \centering
   \includegraphics[angle=0,width=8cm]{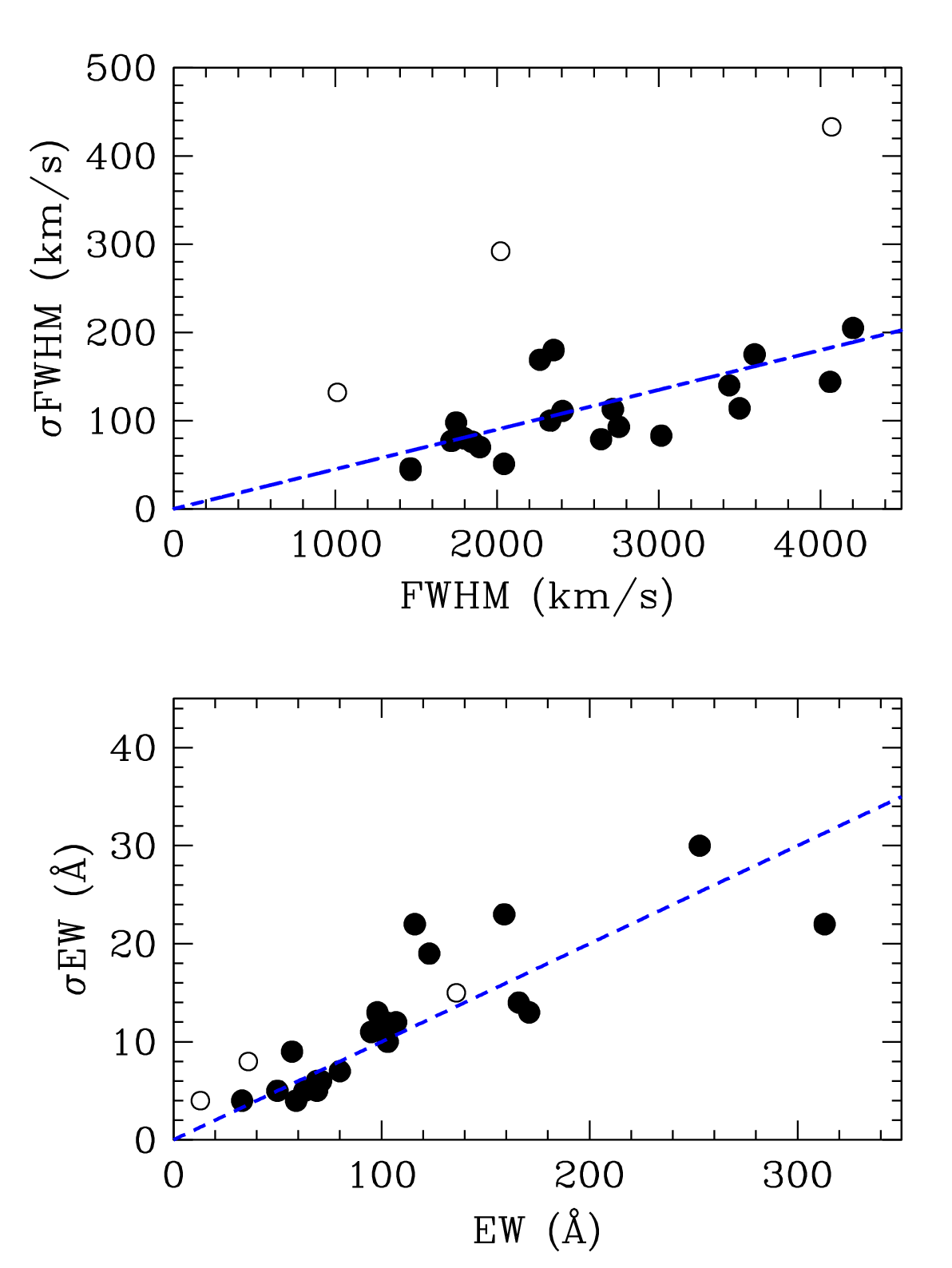}
   \caption{Orbital variability of FWHM (upper panel) and EW (lower panel) for 24 epochs of 
   10 BH XRTs with $\geq$50\%~orbital coverage. Open circles mark the three epochs with S/N$\lesssim$3 spectra
   and thus dominated by statistical noise. The dashed blue lines show linear fits to the data.
   }
              \label{fig:a2}
    \end{figure}

To estimate $\sigma({\rm FWHM})_{\rm orb}$ and  $\sigma({\rm EW})_{\rm orb}$, we focus on 24 epochs of 10 BHs 
with $\geq$50 \% orbital coverage. In principle, one might expect $\sigma({\rm FWHM})_{\rm orb}$ and $\sigma({\rm EW})_{\rm orb}$ 
to depend on geometrical effects, different accretion disc structures or mass accretion rates. However, we see no evidence for a trend in the 
amplitude of the FWHM and EW variability with fundamental parameters such as  binary inclination and $P_{\rm orb}$, 
a proxy for mass transfer rate. Conversely, Fig.~\ref{fig:a2} shows that the amplitude of the variability increases with the 
mean value, with a linear fit giving  $\sigma({\rm FWHM})_{\rm orb}=0.05~{\rm FWHM}$ and $\sigma({\rm EW})_{\rm orb}=0.10~{\rm EW}$. 
Three epochs (the 2022 INT campaign on V404 Cyg, the 1995 WHT on GS2000+25 and the 2013 GTC 
on Swift J1357.2-0933) have significantly larger variability in FWHM than the remaining 21, but this is
caused by statistical noise because the individual spectra in these campaigns   
have very poor signal-to-noise ratios S/N$\lesssim$3 compared to the rest, which typically have S/N$\gtrsim$10. 
In any case, excluding these three epochs does not change the linear fits. 
The effect of these three epochs on the EW variability (which is an integrated line flux) is otherwise negligible, 
despite the poor quality of the spectra. On the basis of Fig.~\ref{fig:a2}, and in the absence of a better approach, we decided to adopt orbital 
fractional errors that are constant for all the systems, i.e. $\sigma({\rm FWHM})_{\rm orb}/{\rm FWHM}=0.05$ and $\sigma({\rm EW})_{\rm orb}/{\rm EW}=0.10$.

Figure~\ref{fig:a1} also depicts clear changes in the FWHM and EW mean values from epoch to epoch. These 
might be caused by geometric changes in a precessing accretion disc (see 
\citealt{zurita02a}, \citealt{torres04}, \citealt{calvelo09}, \citealt{zurita16}),  
although stronger support is needed for confirmation.  
Alternative explanations, such as fluctuations in the mass accretion rate, may also be responsible. For example, 
\cite{cantrell08} has shown that quiescent BH XRTs can sometimes transition between different optical states 
with distinct associated levels of aperiodic variability. In addition, a long-term brightening of the optical continuum 
has been reported in several XRTs, possibly caused by matter accumulating in the disc between outbursts 
(see \citealt{russell18} and included references).     

\begin{figure}
   \centering
   \includegraphics[angle=0,width=8cm]{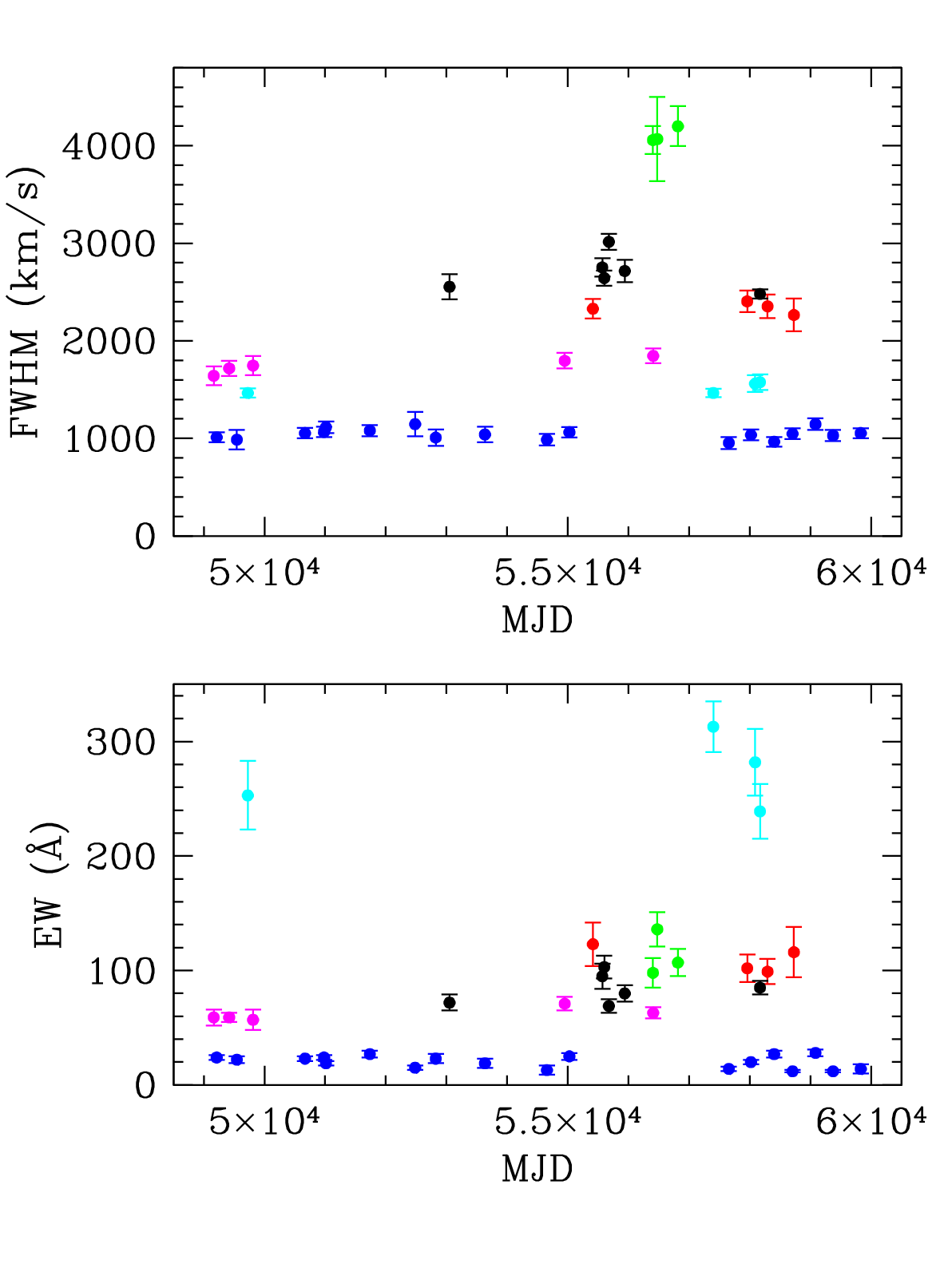}
   \caption{Long-term (secular) variation of FWHM and EW in a sample of quiescent BH XRTs. The plot 
   covers a period of 30 years. Every point represents an orbital average over a single epoch. The color 
   code is as follows: blue (V404 Cyg), cyan (GRO J0422+32), magenta (N. Mus 91), red (XTE J1859+226), 
   black (XTE J1118+480) and green (SWIFT J1357.2-0933).}
              \label{fig:a3}
    \end{figure}

To quantify the long-term variations in FWHM and EW we have 
calculated the standard deviation of orbitally averaged measurements in 9 BHs with $\geq$25 \% phase coverage 
spanning  at least 3 different epochs. Figure~\ref{fig:a3}, for example, shows the secular FWHM and EW variability of a sample 
of BH XRTs through different epochs. 
The small sample size does not allow a detailed analysis, although the plot suggests that there is a stable mean value  
for each system with superimposed secular variability. The amplitude of the variability also appears to increase with the mean, 
so we again assume that a constant fractional variability can be adopted for all the systems. 
The average fractional standard deviation of the 9 BHs is  $\sigma({\rm FWHM})_{\rm sec}= 
0.04~{\rm  FWHM}$ and $\sigma({\rm EW})_{\rm sec}=0.13~ {\rm EW}$. We consider these as a new source of systematic 
error that need to be added quadratically to FWHM and EW measurements obtained from single-epoch data (e.g. N. Vel 93).

The FWHM and EW values given for each epoch in Table~\ref{table:a1} do include the above systematic orbital 
and secular uncertainties, where necessary. With these new added systematic uncertainties, we have calculated the 
mean and standard deviation in the distribution of individual FWHM and EW values for each of the 20 BH XRTs. 
These are the final numbers listed in Table~\ref{table:fwhm_ew} and used in the main body of the paper.

\onecolumn

\begin{TableNotes}
\scriptsize\item
\textbf{Notes. }
\tablefoottext{a}{Values derived from a single or phase-averaged spectrum.}
\tablefoottext{b}{Values derived from a digitised phase-averaged spectrum.}
\tablefoottext{c}{Following (17) we have corrected the EW by a 27 per cent contribution of the interloper to the continuum.}
\tablefoottext{d}{EW values have been digitised from a figure of the corresponding paper.}
\end{TableNotes}

{
\setlength{\tabcolsep}{2.pt}\setlength{\LTcapwidth}{\textwidth}\scriptsize\centering

\begin{longtable}{l c c l c c c c c c}
\caption{\label{table:a1} Multi-epoch FWHM and EW of \ha~lines in quiescent BH XRTs}\\
\hline\hline\\
[-0.15cm] 
Object & Telescope & Year & Day & MJD  & \# spectra & Orbital  & FWHM  & EW & References \\
& & & & & &  Coverage & (\kms) &  (\AA)  & \\
\endfirsthead
\caption{continued.}\\
\hline\hline\\
[-0.15cm] 
Object & Telescope & Year & Day & MJD   & spectra & Orbital  & FWHM  & EW & References \\
& & & & & (\#) &  Coverage & (\kms) &  (\AA)  & \\
\hline\\[-0.15cm]
\endhead
\hline\insertTableNotes\endfoot\hline\\\endlastfoot\hline\\[-0.15cm]

\hline
 V404 Cyg            &   WHT         & 1993 & 12 Aug                                   &  49212.58 &  6 &  5 \%  & 1011$\pm$52   & 24$\pm$2  & 1  \\
                             &   WHT        &  1994 & 6 \& 13-14 July                      &  49544.50 & 21 & 20 \% &  988$\pm$101 & 22$\pm$3  &  2  \\
                             &   WHT        & 1997 & 6-7 Aug                                   & 50668.16 &  4 & 10 \% & 1054$\pm$53 & 23$\pm$2    &  2  \\
                             &   Keck        &  1998 & 17 June                                 & 50982.50 &  1\tablefootmark{a}  &  5 \%  & 1067$\pm$53   & 24$\pm$2 &  3  \\
                             &   WHT        & 1998 &  6 \& 19-20 July                      & 51014.90 & 40 & 10 \% & 1114$\pm$59 & 19$\pm$2    &  3 \\
                             &   WHT        & 2000 & 10 July                                    & 51736.73 &  2  &   5 \% & 1080$\pm$56 & 27$\pm$3  &  3  \\
                             &   INT          & 2002 & 7 July                                      & 52483.69 &  2 &   5 \% & 1147$\pm$123 & 15$\pm$2   &  3 \\
                             &   INT/WHT & 2003 & 1-5 \& 8 July                            & 52824.80 & 19 & 35 \% & 1009$\pm$84 & 23$\pm$4   &  3  \\
                             &   WHT       & 2005 & 16-22 Sept                              & 53634.25 & 23 & 35 \% & 1040$\pm$80 & 19$\pm$4   &  3  \\
                             &   SPM 2.1 m & 2008 & 5-8 July                                 & 54654.74 &   6 & 15 \% &   987$\pm$58 & 13$\pm$4   &  3  \\
                             &   Keck       & 2009 & 13 July                                    & 55025.98 &  14 & 5 \% &   1063$\pm$53 & 25$\pm$3   &  4  \\  
\hline                          
  \multicolumn{10}{c}{\it After 2015 outburst}\\ 
\hline
 V404 Cyg            &   NOT        & 2016 & 21 Sept                                              & 57653.39 &   2 &  5 \% &   954$\pm$61 & 14$\pm$2   &  5 \\
                             &   NOT       & 2017 & 20 Sept                                               & 58017.58 &   2 &  5 \% &  1036$\pm$56 & 20$\pm$2   &  5  \\
                             &   NOT      & 2018 & 8 Aug \& 7 Nov                                    & 58400.71 &   4 &  5 \% &   966$\pm$49 & 27$\pm$3   &  5  \\
                             &   NOT      & 2019 & 10 June \& 18 Oct                                & 58704.16 &   4 &  10 \% &  1049$\pm$55 & 12$\pm$1   &  6  \\
                             &   NOT      & 2020 & 18 Aug                                                  & 59080.44 &   2 &  5 \% &  1146$\pm$59 & 28$\pm$3   &  6  \\
                             &   NOT      & 2021 & 21 June                                                & 59370.68 &   2 &  5 \% &  1030$\pm$58 & 12$\pm$1   &  6  \\
                             &   INT      & 2022 & 12-18 Aug, 3,5,13-14 Sept \& 11 Nov   & 59827.31 &  92 & 70 \% &  1013$\pm$132 & 13$\pm$4   &  6  \\
\hline
\hline
BW Cir                 &   VLT  & 1995  &  3-4 Apr \& 3-4 June                               &  49864.19  &  18 &  30 \% &  1096$\pm$84 & 53$\pm$7 & 3  \\      
                            &   VLT  & 1996  &  24-25 March \& 22-23 May                   & 50169.11  &  14 &  35 \% &  1083$\pm$85 & 54$\pm$10 & 3 \\      
\hline                          
  \multicolumn{10}{c}{\it After 1997 outburst}\\ 
\hline
BW Cir                 &   VLT  & 2000  &  22-23 Aug                                        &  51780.16  &   2 &  10 \% &    979$\pm$53 & 53$\pm$5 & 7  \\      
                            &   VLT  & 2003  &  22-23 June                                       &  52814.46  &  13 &  15 \% &  1204$\pm$37 & 47$\pm$3 & 7  \\      
                            &   VLT  & 2004  &  14-15 \& 25-27 Apr                          &  53142.76  &   42 &  40 \% &    990$\pm$37 & 60$\pm$3 & 7  \\      
                            &   VLT  & 2006  &  27 Feb,  9,18 \& 20-22 March          &  53814.40  &   11 &  25 \% &   1053$\pm$84 & 52$\pm$10 & 8  \\      
\hline
\hline            
GX 339-4            &  VLT  & 2016  & 22 May, 9-10, 12, 14, 30-31 Aug \& 3-7 Sept & 57631.90 & 16\tablefootmark{b}  & 45 \% & 855$\pm$52  & 76$\pm$14 &  9  \\      
\hline
\hline
XTE J1550-564        &  VLT &  2001  & 24-27 May                                     & 52053.78  & 18  & 30 \% & 1491$\pm$99 & 29$\pm$4 & 10 \\      
\hline                          
  \multicolumn{10}{c}{\it After 2003 outburst}\\ 
\hline
XTE J1550-564        &  Magellan Clay &  2008  & 28 June \& 3-4 Aug       & 54645.56  & 16  & 30 \% & 1419$\pm$139 & 29$\pm$6 & 11 \\      
\hline
\hline
MAXI J1820+070       & GTC  & 2023  & 18-19, 23 July \& 15, 17 Aug  & 60153.97 & 11 & 60 \%& 1678$\pm$85 & 75$\pm$19 & 6 \\      
\hline
\hline
N. Oph 77             &  Victor M. Blanco          & 1993  & 23 May &  49131.09 & 1\tablefootmark{a}  & 65 \%  & 1758$\pm$88 & 80$\pm$8 & 12 \\      
                             &  Victor M. Blanco          & 1994  & 4-6 July &  49539.87 & 1\tablefootmark{b}  & 65 \%  & 2053$\pm$109 & 46$\pm$5  & 12 \\      
                             &  Keck                             & 1996  & 12 May & 50215.98 & 1\tablefootmark{b}  & 20 \%  & 1764$\pm$91& 79$\pm$8 & 13 \\      
                             &  Keck                             & 1996  & 14 July & 50278.87 & 1\tablefootmark{b}  & 50 \%  & 2001$\pm$107& 22$\pm$2 & 13 \\      
\hline
\hline
N. Mus 91             & AAT                             & 1993 &  25 June                            & 49164.89  & 2 & 10 \%  &  1642$\pm$96  & 59$\pm$7 & 14  \\      
                             & NTT                             & 1994 &  5-7 March                         & 49419.13  & 13 & 90 \%  &  1718$\pm$77  & 59$\pm$4 & 14  \\      
                             & NTT                             & 1995 &  3-4 April                            & 49812.21  & 16 & 95 \%  &  1747$\pm$98  & 57$\pm$9 & 14  \\      
                             & Magellan Clay             & 2009 &  25 April                             & 54947.47  & 40 & 90 \%  &  1796$\pm$80  & 71$\pm$6 & 15  \\      
                             & VLT                             & 2013 &  12-14, 28 April \& 9 May   & 56407.04  & 17 & 80 \%  &  1847$\pm$76  & 63$\pm$5 & 16  \\      
\hline
\hline
MAXI J1305-704    &  VLT                         & 2016  &  31 March        & 57479.70  & 16 & 95 \%  & 2350$\pm$203 & 33$\pm$6\tablefootmark{c} & 17 \\      
\hline
\hline
GS 2000+25         &   Keck                     & 1995  & 22 July                               &   49920.81 & 1\tablefootmark{a}   & 100 \% & 2261$\pm$117 & 24$\pm$3 & 18 \\      
                             &   WHT                     & 1995  & 24-26 July                          &   49925.29 & 1\tablefootmark{a}   & 100 \% & 2119$\pm$121 & 34$\pm$4  & 19 \\     
\hline
\hline
A 0620-00            &   WHT                     &  1991-92     & 31 Dec-1 Jan            & 48621.67  & 1\tablefootmark{b} &   90 \% & 1921$\pm$96 & 61$\pm$7\tablefootmark{d}  &  20  \\      
                            &   VLT                       &  2000           & 6, 16 \& 20 Dec        & 51889.72  & 20 &   40 \% & 1939$\pm$69  & 44$\pm$12 & 21 \\
                            &   Magellan Clay      &  2006           & 14-16  Dec                & 54085.67  & 1\tablefootmark{b} & 100 \% & 1855$\pm$93  & 74$\pm$7\tablefootmark{d} & 22 \\
                            &   GTC                     &  2012           & 5-6 Dec                     & 56268.18  & 40 &   55 \% & 2043$\pm$51  & 50$\pm$5  & 23 \\
                            &   GTC                     &  2013           & 7 Jan                         & 56300.49  & 38 &   80 \% & 1894$\pm$70  & 69$\pm$5  & 23 \\
                            &   GTC                     &  2018           & 17 Feb                       & 58167.39  & 12 &    5 \% & 1776$\pm$92  & 61$\pm$7  & 24 \\
\hline
\hline
XTE J1650-500   &  VLT                       &  2002           & 10 June                      & 52436.70 & 1\tablefootmark{a}  &  100 \% &1898$\pm$121 & 9$\pm$2 & 25-26 \\      
\hline
\hline
N. Vel 93             &   Keck                    & 1998-99       & 1 Feb, 5-6 Mar 98 \& 22 Jan 99  & 52436.70 & 1\tablefootmark{a}  & 70 \% & 2055$\pm$124 & 63$\pm$10 & 27 \\      
\hline
\hline
N. Oph 93           & GTC.                     &  2020           & 22 June                      & 59023.53  & 1\tablefootmark{a}   & 30 \% & 2267$\pm$194 & 89$\pm$15 & 28 \\      
\hline
\hline
XTE J1859+226  & GTC                     & 2010            & 17 June \& 13 Aug      & 55413.55  & 10 &  80 \%  & 2329$\pm$100 & 123$\pm$19& 29 \\      
                            & GTC                     & 2017            & 22-23 July                   & 57957.88  & 26 & 100 \% & 2405$\pm$111 & 102$\pm$12& 30 \\
                            & GTC                     & 2018            & 20 June                       & 58290.62  &  2  & 15 \% & 2353$\pm$120    &  99$\pm$11& 24 \\
                            & GTC                     & 2019            & 22-29 \& 30-31 Aug     & 58725.22  & 10  & 85 \% & 2265$\pm$169  &  116$\pm$22& 6 \\                                  
\hline
\hline
KY TrA               &  VLT                       & 2016           &  4 \& 7 April                  & 57484.06  & 6 & 30 \% &  2167$\pm$185 & 56$\pm$18 & 31 \\      
\hline
\hline
GRO J0422+32   &  WHT                   & 1994-95     &  28 Dec 94, 1 \& 22 Feb 95  & 49726.18 & 17 & 100 \% & 1465$\pm$46 & 253$\pm$30 & 32 \\      
                            &  GTC                   & 2016           &  9 Jan                                   & 57397.49 & 18 & 100 \% & 1466$\pm$44 & 313$\pm$22 & 23 \\      
                            &  GTC                   & 2017          &  28 Nov                                  & 58086.50 &  3 &   20 \% & 1562$\pm$86 & 282$\pm$29  &  6 \\      
                            &  GTC                   & 2018           &  17 Feb                                 & 58167.36 &  2 &  15 \% & 1577$\pm$81 & 239$\pm$24  & 24 \\      
\hline
\hline
XTE J1118+480   &  Keck                  &  2004          & 14 Feb                                  & 53050.00 & 1\tablefootmark{a} & 100 \%  &  2555$\pm$128  & 72$\pm$7 & 33-34 \\      
\hline
  \multicolumn{10}{c}{\it After 2005 outburst}\\ 
\hline
XTE J1118+480   &  GTC                  &  2011          & 6 Jan                                     & 55568.66 & 25 &  70 \%  &  2753$\pm$93 & 95$\pm$11 & 35-36 \\      
                            &  GTC                  &  2011          & 7 Feb                                     & 55600.63 & 36 &  95 \%  &  2643$\pm$79 &103$\pm$10 & 35-36 \\      
                            &  GTC                  &  2011          & 24 Apr                                    & 55676.53 & 36 & 100 \%  &  3015$\pm$83 & 69$\pm$6 & 35-36 \\    
                            &  GTC                  &  2012          & 11 Jan                                   & 55938.68 & 34 &  95 \%  &  2716$\pm$113 & 80$\pm$7 & 36-37 \\    
                            &  GTC                  &  2018          & 17 Feb                                   & 58167.47 &  9 &  35 \%  &  2480$\pm$48 & 85$\pm$6 & 24 \\                              
\hline
\hline
Swift J1753.5-0127   &  GTC           &  2018          & 12 July                                  &  58312.50 & 14 &  95 \% & 3592$\pm$175 & 166$\pm$14 & 38 \\      
                                 &  GTC           &  2018          & 13 July                                   &  58313.49 & 14 & 100 \% & 3435$\pm$140 & 159$\pm$23 & 38 \\      
                                 &  GTC           &  2018          & 14 July                                   &  58314.49 & 12 & 100 \% & 3499$\pm$114 & 171$\pm$13 & 38 \\      
\hline
\hline
Swift J1357.2-0933   &  VLT           &  2013          &  13, 17 Apr \& 3 May             & 56402.02 & 24 &  90 \%  &  4058$\pm$144& 98$\pm$13 & 39 \\      
                                 &  GTC           &  2013          &  29-30 June                          & 56473.99 &  4 &  80 \%  &  4068$\pm$433& 136$\pm$15 & 3 \\  
                                 &  GTC           &  2014          &  29-30 Apr, 2-3 \& 28 June   & 56813.81 & 42 & 100 \%  &  4200$\pm$205 & 107$\pm$12 & 40 \\  
\hline
\hline
MAXI J1659-152      & VLT             & 2013           & 6 June                                   &  56449.79 & 1\tablefootmark{a} & 30 \% & 3309$\pm$361 &130$\pm$23 & 41 \\      
\hline
\hline

\end{longtable}

\tablebib{(1)~\citet{casares94}; (2) \citet{casares96}; (3) \citet{casares15}; (4) \citet{gonzalez11}; 
(5) \citet{casares19}; (6) this paper;  (7) \citet{casares04}; (8) \citet{casares09}; (9) \citet{heida17}; 
(10) \citet{orosz02}; (11) \citet{orosz11}; 
(12) \citet{remillard96};  (13) \citet{filippenko97}; (14) \citet{casares97}; (15) \citet{wu15}; (16) \citet{gonzalez17}; 
(17) \citet{mata21}; (18) \citet{casares95a}; (19) \citet{filippenko95}; (20) \citet{marsh94};  (21) \citet{gonzalez10}; 
(22) \citet{neilsen08}; (23) \citet{casares22};  (24) \citet{casares-torres18}; (25)  \citet{sanchez02}; (26) \citet{orosz04}; 
(27) \citet{filippenko99}; (28) \citet{casares23}; (29) \citet{corral11};  (30) \citet{yanes22}; (31) \citet{yanes24}; 
(32) \citet{casares95b}; (33) \citet{gonzalez06};  (34) \citet{gonzalez08b};  (35) \citet{gonzalez12};  
(36) \citet{zurita16};  (37) \citet{gonzalez14};  (38) \citet{yanes25};  
(39) \citet{torres15};  (40) \citet{mata15}; 
 (41) \citet{torres21}
}
}

   \begin{table*}[b!]
      \caption[]{Quiescent epochs and magnitudes}
         \label{table:a2}
\centering                          
\begin{tabular}{l c c c c}        
\hline\hline                
     Source              &  Outburst   & Quiescence  & Quiescent    & References \\
                              &   (Year)       & (MJD)          &     Mags        &  for  Mags       \\
 \hline
V404 Cyg                & 1989         & 48814-57190 &   V=18.42$\pm$0.02,   R=16.52$\pm$0.01                            &   (1) \\
                                & 2015         & 57405-           &   V=18.2$\pm$0.3                                                                   &   (2) \\
 \hline
BW Cir                   & 1987          & $\sim$47200-50754  &   R$\sim$20.2, I$\sim$19.2                                           &   (3) \\ 
                               & 1997          & $\sim$51135-57000  &  R$\sim$20-21,  i'=19.7-20.8                                       &   (3), (4) \\      
 \hline
GX 339-4 \tablefootmark{a}  & 2014-5  &  57579-57990  &   V$\simeq$20, R$\simeq$19, i$\simeq$18.7               &   (5) \\      
 \hline  
XTE J1550-564       & 2001         & 52020-52719            &  V$\simeq$22                                                                  &   (6) \\      
                                & 2003         & 52764-                      &  V$\simeq$21.9                                                               &   (7) \\      
 \hline
MAXI J1820+070    & 2018         & 60106-                     &  g'=19.31$\pm$0.02, r'=18.58$\pm$0.01, i'=18.22$\pm$0.01 &  (8) \\      
 \hline
N. Oph 77               & 1977          & 48719-                   &  V=21.5$\pm$0.1                                                                    &   (9)  \\      
 \hline
N. Mus 91               & 1991          & 48736-                  &  V=20.66$\pm$0.03, R=19.75$\pm$0.27, I=19.00$\pm$0.22 &   (10)  \\      
 \hline
MAXI J1305-704     & 2012          & 56771-                 &   r'=21.69$\pm$0.17                                                                 &   (11) \\      
 \hline
GS 2000+25           & 1988          & 47763-                 &  R=21.21$\pm$0.15                                                                 &   (12) \\      
 \hline
A 0620-00               &  1975         & 43053-                 & V=18.25, R=17.00                                                                    &  (13) \\      
 \hline
XTE J1650-500 \tablefootmark{b}  &  2001  & $\sim$52490- &  V$\sim$24, R$\sim$22                                                &  (14) \\      
 \hline
N. Vel 93                 &  1993        &  49846-               &  R=20.6$\pm$0.1                                                                       &  (15) \\      
 \hline
N. Oph 93               & 2016        & 58208-                 &  i'=21.39$\pm$0.15                                                                   &  (16) \\      
\hline
XTE J1859+226      & 1999        & 51781-59250      & V=23.39$\pm$0.09,  R =22.48$\pm$0.07                                 &  (17) \\      
\hline
KY TrA \tablefootmark{c}    & 1990      & 48290-      &   V=23.6$\pm$0.1, R=22.3$\pm$0.1, I=21.47$\pm$0.09         &  (18) \\      
\hline
GRO J0422+32       & 1992           & 49600-            & R=20.94$\pm$0.11                                                                    &  (19) \\      
\hline
XTE J1118+480       & 2000          & 52311-53375   & R=18.93$\pm$0.01                                                                     &  (20) \\     
                                & 2005           & 53447-            & R$\sim$19                                                                                   &  (21) \\     
\hline
Swift J1753.5-0127  & 2005           & 57907-60216 & V=22.17$\pm$0.25, I'=21.00$\pm$0.14                                      &  (22) \\      
\hline
Swift J1357.2-0933  & 2011           & 56042-57845  & g'=22.26$\pm$0.38, r'=21.54$\pm$0.35, i'=21.21$\pm$0.36; I$\sim$20-21  &  (23), (24) \\      
\hline
MAXI J1659-152      & 2010           & 56087-            & r'=24.20$\pm$0.08, I=23.3$\pm$0.1                                           &  (25) \\      
                        \noalign{\smallskip}
            \hline
 \end{tabular}
\tablebib{(1)~\citet{casares93}; (2) \citet{casares22}; (3) \citet{casares09}; (4) \citet{koljonen16};
(5) \citet{russell17}; (6) \citet{orosz02}; (7) \citet{orosz11}; (8) \citet{baglio23}; (9) \citet{remillard96}; 
(10) \citet{king96b};(11) \citet{mata21}; (12) \citet{callanan91}; (13) \citet{ciatti77};  (14) \citet{garcia02}; 
(15) \citet{shahbaz96};  (16) \citet{saikia22}; (17) \citet{zurita02b}; 
(18) \citet{zurita15};   (19) \citet{garcia96};  (20) \citet{zurita02c}; (21) \citet{zurita06}; 
(22) \citet{zhang17}; (23) \citet{shahbaz13};  (24) \citet{russell18};  (25) \citet{corral18}.
}
\tablefoot{\\
\tablefoottext{a}{It shows outbursts every $\sim$2.5 years. The following outburst took place in 2017.}
\tablefoottext{b}{The quiescent R-band magnitude quoted by \cite{garcia02} is consistent with our own PSF analysis of  a  
DECaPS image taken on 29 April 2017, which yields r'=22.10$\pm$0.12,}  
\tablefoottext{c}{The optical counterpart of KY TrA was lost after the V$>$21 upper limit reported by \cite{murdin77} during the decay of the 1974 discovery outburst. The 
 quiescent counterpart was only recovered by \citet{zurita15} 17 years after a new mini-outburst was serendipitously discovered in 1990.}    
}
\end{table*}

\twocolumn

\section{ Empirical $P_{\rm orb}-T_{\rm eff}$ relation for low-mass donors in BH X-ray transients}
\label{ap:teff_porb}

In order to derive an empirical $P_{\rm orb}-T_{\rm eff}$ relation for donor stars in BH XRTs  we start by collecting 
spectral types from the literature. The compilation is divided into two categories: low-mass companions 
$M_{2}\lesssim$1.5 \msun~ and  intermediate-mass companions $M_{2}\approx2-5$ \msun. The motivation for 
doing so is two-fold. On the one hand, XRTs with intermediate-mass companions (IMXBs) are thought to be 
precursors of those with low-mass companions (LMXBs), and thus  represent a different evolutionary phase 
\citep{podsiadlowski03}. On the other, high luminosity donors in IMXBs totally veil disc emission lines,  such as \ha, 
and, hence, are of little interest in the context of this paper. We therefore focus here on the group of BH LMXBs 
(Table~\ref{table:b1}),  although information on BH IMXBs is also provided for completeness (Table~\ref{table:b2}). 
The categorisation of BW Cir is somewhat controversial because its donor is an early G-type star with a mass of 
$\approx1-2.4$ \msun~that sits within the Hertzsprung gap \citep{casares04}. However, because of its moderately 
strong \ha~line, we tentatively consider BW Cir as a BH LMXB in this work. 
 
Reported spectral types have been determined following three different methods: (1) a qualitative classification, 
based on the presence (or lack thereof) of distinct spectral features (e.g. the $G$-band at $\sim$4310 \AA, 
the Mg Ib triplet at $\sim$5170 \AA, TiO or CO molecular bands, etc.), either in the visible (VIS SPEC) or 
near-infrared (NIR SPEC) part of the spectrum. In some cases, the classification is more quantitative as it 
relies on figures of merit such as the intensity of the highest cross-correlation peak or a $\chi^2$ minimisation 
of residuals after template subtraction.  (2) Model fits to the observed (donor's dominated) multi-wavelength 
spectral energy distribution (SED). (3) a direct $T_{\rm eff}$ determination derived by fitting libraries of 
synthetic stellar models (SPEC FIT). 

In the case of VIS/NIR SPEC and SED methods we transformed spectral types into $T_{\rm eff}$ values. To do so 
we assume that LMXB donors below the {\it bifurcation period}  ($P_{\rm orb}\lesssim$18 h; see \citealt{podsiadlowski02}) 
can be approximated by main sequence (MS) stars and, thus, adopt the $T_{\rm eff}$ scale from the empirical collection 
of table 5 in \citet{pecaut13}. For long period LMXBs ($P_{\rm orb}\gtrsim$3 d) we follow instead the $T_{\rm eff}$ 
scale of giant stars from \citet{vanbelle99}. Finally, LMXB donors with 18 h $\lesssim P_{\rm orb}\lesssim$ 3 d 
are considered subgiants, and  their  $T_{\rm eff}$ values interpolated between the MS and giant scales of 
\citet{pecaut13} and \citet{vanbelle99}. 

To verify the reliability of these assumptions we have compared average densities $\langle \rho \rangle$ of LMXB 
companions with those of MS and giants stars of similar spectral types. The density of a Roche-lobe filling star 
is obtained by bringing the Roche lobe geometry into Kepler's third law. Adopting Paczy\'nski's approximation 
for the volume-averaged Roche-lobe radius \citep{paczynski71} allows cancelling out the dependence on binary 
mass ratio $q$, leading to $\langle \rho \rangle \approx110\times P_{\rm orb}^{-2}$, where $P_{\rm orb}$ is given 
in units of hours and $\langle \rho \rangle$ in gr cm$^{-3}$ \citep{frank02}. If instead we use Eggleton's more 
accurate approximation \citep{eggleton83}, we find  

 \begin{equation}
\langle \rho \rangle = 92.6 \times \frac{\left[0.6~q^{2/3} + \ln \left( 1+q^{1/3} \right)  \right]^{3} }{q~\left( 1+q \right) }\times P_{\rm orb}^{-2} ,\,
\label{eq:rho}
 \end{equation}
 
\noindent
equation that weakly depends on $q$. We note the former expression $\langle \rho \rangle \approx110\times P_{\rm orb}^{-2}$ is 
recovered for $q=0.22$, which seems apropriate for cataclysmic variables. BH LMXBs, on the other hand, have more extreme mass 
ratios, with typical values $q\simeq0.06$ (cf \citealt{casares16}), leading to $\langle \rho \rangle \approx186\times P_{\rm orb}^{-2}$.  
Table~\ref{table:b3} presents an up-to-date collection of mass ratios in BH LMXBs and implied donor densities, according to 
eq.~\ref{eq:rho}. In any case, it should be noted that the use of Paczy\'nski's approximation is equally valid as it leads to tiny 
differences in density on the order of $\approx$2-4 \%.  A direct comparison with MS and giant star densities supports our choice 
of  $T_{\rm eff}$ scales for LMXB donors i.e. average donor densities above $P_{\rm orb}\gtrsim$3 d are in the range of giant stars  
while those with $P_{\rm orb}\lesssim$18 h are consistent with typical MS or slightly evolved (i.e. oversized) stars.   
 
\begin{figure} 
  \centering
    \includegraphics[angle=-90,width=\columnwidth]{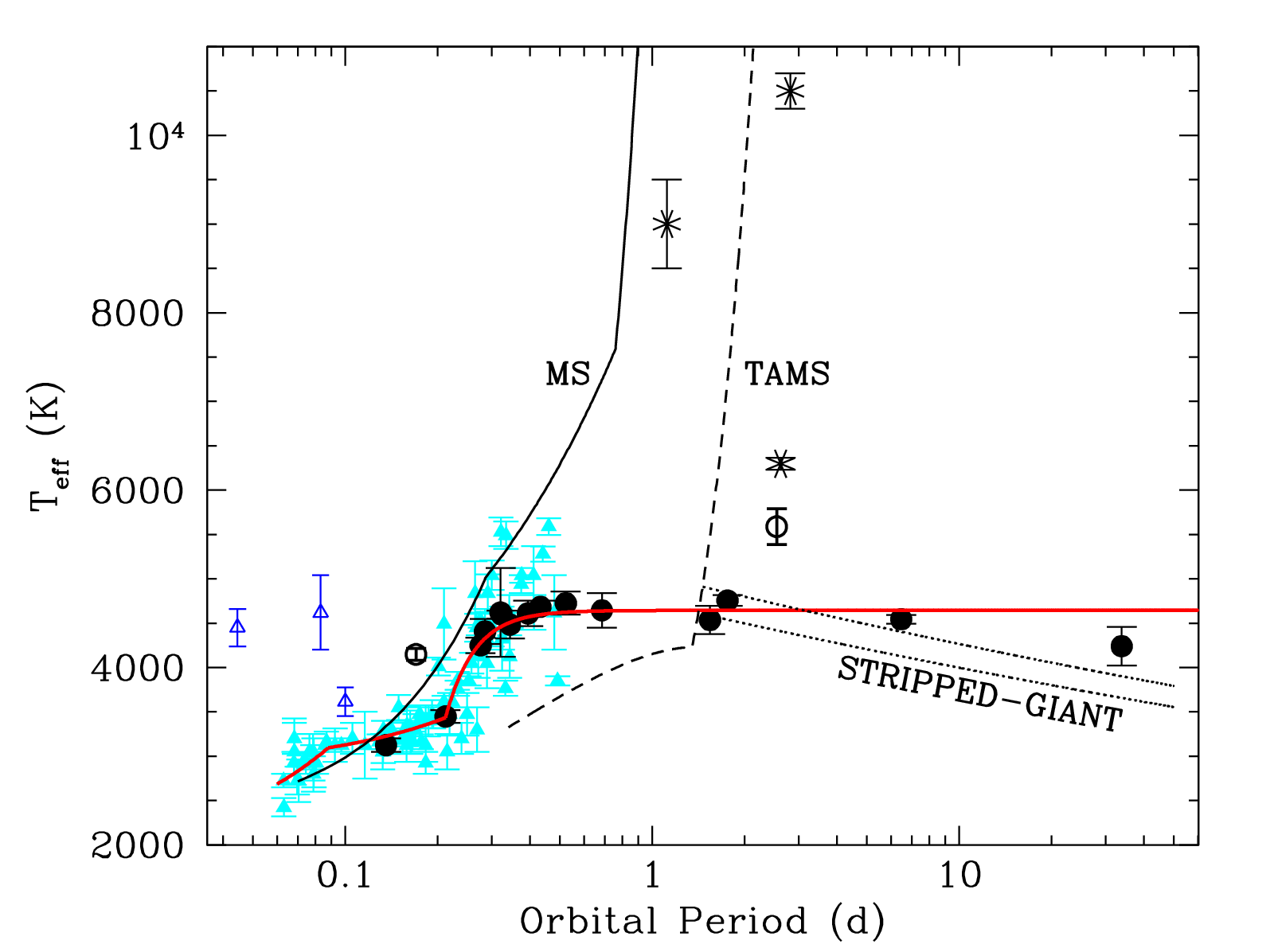}
      \caption{Observed $T_{\rm eff}$ vs orbital period ($P_{\rm orb}$) for donor stars in BH XRTs. Asterisks indicate 
      $\approx2-5$ \msun~donor stars (IMXBs) while solid black circles low-mass  $\lesssim1.5$ \msun~donors (LMXBs). 
      The open circles mark the position of the early G-type donor in BW Cir and the nuclear-evolved companion in 
      XTE J1118+480.  Solid triangles in cyan indicate donor stars in CVs from the compilation of \cite{knigge06}. 
      The blue open triangles indicate the three nuclear-evolved donors in EI Psc, QZ Ser and SDSS 1702+3229. 
      The red line shows our empirical fit to the group of CVs and BH LMXBs, excluding BW Cir and the 
      nuclear-evolved companions in XTE J1118+480 and the three CVs. For comparison, we also plot 
      $T_{\rm eff}$ tracks of main sequence (MS), terminal age main sequence (TAMS) 
      and stripped-giant stars that would fill the Roche lobes for a given $P_{\rm orb}$.
               }
         \label{ap:porb_teff}
   \end{figure}

Tables~\ref{table:b1} and \ref{table:b2} summarise the spectral types collected from the literature and their associated $T_{\rm eff}$ 
numbers. When several $T_{\rm eff}$  values are present for a given system, the average (marked in bold) is selected. This was 
computed as the unweighted mean after randomising the individual (independent)  measurements. Here we have assumed flat 
probability distributions for all methods except SPEC FIT, where a normal distribution is adopted. The evolution of $T_{\rm eff}$ 
with $P_{\rm orb}$ is presented in Fig.~\ref{ap:porb_teff}. For reference, we also plot $P_{\rm orb}-T_{\rm eff}$ tracks for MS 
and terminal age main sequence (TAMS) stars that would fit in their corresponding Roche lobes. To do so, we used 
eq.~\ref{eq:rho} with $q=0.06$ and mass-radius and mass-$T_{\rm eff}$ relations for MS \citep{pecaut13} and TAMS 
 \citep{bertelli08,bertelli09}. In addition, we depict stripped-giant models to track the location of low-mass donors that 
 have evolved off the MS. The radius and luminosity (and thus $T_{\rm eff}$) of stripped-giant stars are uniquely determined 
 by the mass of the degenerate helium core $m_{\rm c}$, which is constrained between the total stellar mass $M_2$ and the 
 Sch\"onberg-Chandrasekhar limit $\sim 0.17 M_2$ \citep{webbink83, king93}. Both models, $m_{\rm c}=0.17 M_2$ (lower line) 
 and $m_{\rm c}=M_2$ (upper line), are displayed.

\begin{table*}
\caption{Spectral types and $T_{\rm eff}$ of donor stars in  BH LMXBs}             
\label{table:b1}      
\centering          
\begin{tabular}{l l l c l c}     
\hline\hline       
Object & $P_{\rm orb}$  & Spect. Type & Method & $T_{\rm eff}$  & References for \\
          &          (d)          &                     &              & ($^{\circ}$K) &      Spect. Type   \\
\hline
 GRS 1915+105  & 33.85(16) & K1-5 III    & NIR SPEC &  4241$\pm$219 &  1 \\
 \hline
 V404 Cyg            &   6.471170(2) & G9-K1 III & VIS SPEC  & 4570$\pm$110  & 2  \\
                             &             & G8-K2 III &  SED          & 4570$\pm$219  & 3  \\
                             &             & K2-4 III    & NIR SPEC & 4242$\pm$110  & 4  \\
                             &             &    --           & SPEC FIT & 4800$\pm$100  & 5  \\
                             &             &    --           &      --         & {\bf 4544$\pm$48} & --  \\
\hline 
BW Cir                  &   2.54451(8) &  G0-5 IV  & VIS SPEC & 5589$\pm$201  & 6  \\      
\hline            
GX 339-4              &   1.7587(5) &  K1-2 IV  & NIR SPEC & 4756$\pm$60  & 7  \\      
\hline
XTE J1550-564     &   1.5420333(24) &  K2-4 IV  & VIS SPEC & 4536$\pm$160  & 8  \\      
\hline
MAXI J1820+070   &   0.68549(1) &  K3-5 V  & VIS SPEC & 4645$\pm$195  & 9  \\      
\hline
N. Oph 77               &   0.5228(44) &  K1-5 V  & VIS SPEC & 4810$\pm$360  & 10  \\      
                               &             & K3-5 V \tablefootmark{a}     & VIS SPEC & 4645$\pm$195  & 11  \\
                               &             &    --           &      --         & {\bf 4725$\pm$130} & --  \\
\hline
N. Mus 91               &   0.43260249(9) &  K0-4 V  & VIS SPEC & 4950$\pm$330  & 12  \\      
                               &              & K3-5 V   & VIS SPEC & 4645$\pm$195  & 13  \\
                               &              & K3-4 V   & VIS SPEC & 4730$\pm$110  & 14  \\
                               &              & K4-6 V   & VIS SPEC & 4410$\pm$210  & 15  \\
                               &             &    --          &      --         & {\bf 4685$\pm$70} & --  \\
\hline
MAXI J1305-704     &   0.394(4) &    --         & SPEC FIT & 4610$\pm$145  & 16  \\      
\hline
GS 2000+25            &   0.3440915(9) &  K3-7 V  & VIS SPEC & 4445$\pm$395  & 17  \\      
                                &              & K3-6 V   & VIS SPEC & 4520$\pm$320  & 18  \\
                                &             &    --          &      --         & {\bf 4485$\pm$158} & --  \\
\hline
A 0620-00                &   0.32301415(7) &  K5-7 V  & VIS SPEC & 4250$\pm$200  & 19 \\      
                                &              & K4-5 V   & VIS SPEC & 4535$\pm$85   & 20  \\
                                &              &  K3-7 V  & VIS SPEC & 4445$\pm$395  & 21  \\      
                                &              & K3-4 V   & VIS SPEC & 4730$\pm$110  & 22  \\
                                &              & K0-5 V   & NIR SPEC & 4865$\pm$415  & 23  \\
                                &              &  K3-5 V\tablefootmark{b}   & SED  & 4600$\pm$200  & 24  \\    
                                &              &     --         & SPEC FIT & 4900$\pm$100  & 25  \\
                                &              &  K5-7 V   & NIR SPEC & 4250$\pm$200  & 26  \\      
                                &              &  K4-6 V\tablefootmark{c}   & NIR SPEC  & 4410$\pm$210  & 27  \\    
                                &              &     --         & SPEC FIT & 5000$\pm$100  & 28  \\
                                &              &    --          &      --         & {\bf 4598$\pm$45} & --  \\
\hline
XTE J1650-500       &   0.3205(7)  &  K1-7 V\tablefootmark{d}   & VIS SPEC & 4620$\pm$500  & 29  \\      
\hline
N. Vel 93                 &   0.285206(1) &  K0-5 V  & VIS SPEC & 486$\pm$415  & 30  \\      
                                &              & K7-M0 V  & VIS SPEC & 3950$\pm$100  & 31  \\
                                &             &    --          &      --         & {\bf 4407$\pm$142} & --  \\
\hline
XTE J1859+226      &   0.276(3) &  K5-7 V  & VIS SPEC & 4250$\pm$200  & 32  \\      
                                &              & K5-7 V  & VIS SPEC & 4250$\pm$200  & 33  \\
                                &             &    --          &      --         & {\bf 4250$\pm$88} & --  \\
\hline
GRO J0422+32       &   0.2121600(2) &  M0-5 V  & VIS SPEC & 3450$\pm$400  & 34  \\      
                                &              & M0-4 V  & VIS SPEC & 3520$\pm$325  & 35  \\
                                &              & M1-4 V  & VIS SPEC & 3440$\pm$240  & 36  \\
                                &              & M4-5 V  & VIS SPEC & 3125$\pm$75  & 37  \\
                                &              & M0-2 V  & SED          & 3700$\pm$150  & 38  \\
                                &              &    --          &      --         & {\bf 3448$\pm$70} & --  \\
\hline
XTE J1118+480       &   0.16993404(5) &  K7-M0 V  & VIS SPEC & 3950$\pm$100  & 39  \\      
                                &              & K5-M1 V  & VIS SPEC & 4065$\pm$385  & 40  \\
                                &              & K5-9 V\tablefootmark{e} & SED & 4165$\pm$285  & 41  \\
                                &              &     --         & SPEC FIT & 4700$\pm$100  & 42 \\
                                &              & K7-M1 V  & NIR SPEC & 3865$\pm$185  & 43  \\
                                &              &    --          &      --         & {\bf 4148$\pm$66} & --  \\
\hline
SWIFT J1753.5-0127 &   0.1358(8) &  M4-5 V  & VIS SPEC & 3125$\pm$75  & 44  \\      
\hline
\end{tabular}
\tablebib{(1)~\citet{harlaftis04}; (2) \citet{casares94}; (3) \citet{hynes09}; (4) \citet{khargharia10};
(5) \citet{gonzalez11}; (6) \citet{casares04}; (7) \citet{heida17}; (8) \citet{orosz11}; (9) \citet{torres19}; 
(10) \citet{remillard96}; (11) \citet{harlaftis97}; (12) \citet{remillard92}; (13) \citet{orosz96}; (14) \citet{casares97};
(15) \citet{wu15}; (16) \citet{mata21}; (17) \citet{casares95a}; (18) \citet{harlaftis96}; (19) \citet{oke77}; 
(20) \citet{murdin80}; (21) \citet{mcclintock86}; (22) \citet{marsh94}; (23) \citet{shahbaz99a};  (24) \citet{gelino01}; 
(25) \citet{gonzalez04}; (26) \citet{froning07}; (27) \citet{harrison07}; (28) \citet{zheng22};  (29) \citet{orosz04};  
(30) \citet{shahbaz96}; (31) \citet{filippenko99}; (32) \citet{corral11};  (33) \citet{yanes22}; (34) \citet{orosz95}; 
(35) \citet{casares95b}; (36) \citet{harlaftis99}; (37) \citet{webb00}; (38) \citet{gelino03}; (39) \citet{wagner01}; 
(40) \citet{mcclintock01};  (41) \citet{gelino06};  (42) \citet{gonzalez08b};  (43) \citet{khargharia13}; (44) \citet{yanes25}.
}
\tablefoot{\\
\tablefoottext{a}{Although (11) report K3-M0, the clear lack of TiO bands in their spectra strongly suggests $\lesssim$K5.}
\tablefoottext{b}{We adopt an uncertainty of $\pm$1 sub-types on the spectral classification favoured by (24).}
\tablefoottext{c}{We adopt an uncertainty of $\pm$1 sub-types on the spectral classification favoured by (26).}
\tablefoottext{d}{The spectral type is poorly constrained by (28) because of limited quality spectra and thus we adopt a conservative  uncertainty of $\pm$3 sub-types.}
\tablefoottext{e}{Following (26) we adopt an uncertainty of $\pm$2 sub-types on the favoured spectral classification.}
}
\end{table*}

\begin{table*}
\caption{Spectral types and $T_{\rm eff}$ of donor stars in BH IMXBs}             
\label{table:b2}      
\centering          
\begin{tabular}{l r l c l c}  
\hline\hline       
Object & $P_{\rm orb}$  & Spect. Type & Method & $T_{\rm eff}$  & References for \\
          &          (d)          &                     &              & ($^{\circ}$K) &      Spect. Type   \\
\hline
SAX J1819.3-2525 &  2.81730(1) & B9 III  & SPEC FIT & 10500$\pm$200 & 44 \\ 
\hline
GRO J1655-40       &  2.621928(4) & F3-6 IV     & VIS SPEC & 6395$\pm$177 & 45 \\ 
                               &             & F5-7 IV    & VIS SPEC & 6236$\pm$122  & 46  \\
                               &             &     --         & SPEC FIT & 6400$\pm$250  & 47  \\
                               &             &     --         & SPEC FIT & 6500$\pm$50  & 48  \\
                               &             & F5-G0 IV &  SED       & 6150$\pm$350  & 49  \\
                               &             &     --         & SPEC FIT & 6100$\pm$200  & 50  \\
                                &             &    --          &      --         & {\bf 6297$\pm$68} & --  \\
\hline
4U 1543-475           &  1.116407(3) & A2$\pm$1 V & SPEC FIT & 9000$\pm$500 & 51 \\ 
\hline                  
\end{tabular}
\tablebib{(44) \citet{orosz01}; (45) \citet{orosz97}; (46) \citet{shahbaz99b}; (47) \citet{israelian99}; (48) \citet{buxton01}; 
(49) \citet{beer02}; (50) \citet{gonzalez08a}; (51) \citet{orosz98}.
}
\end{table*}

\begin{table*}[h]
 \caption[]{\label{table:a4}Donor star densities in BH LMXBs (marked in bold), compared to those of MS and Giants with same spectral type.}
\label{table:b3}  
\centering    
\begin{tabular}{lllccccc}
 \hline \hline
  Object & $P_{\rm orb}$ & $q$   & $\langle \rho \rangle$ & Spec Type \tablefootmark{a}  &  
  $\langle \rho_{\rm MS} \rangle$ \tablefootmark{b}  & $\langle \rho_{\rm GIANT} \rangle$ \tablefootmark{c} & References  \\
   & (d) &   & (gr cm$^{-3}$) &  & (gr cm$^{-3}$)  & (gr cm$^{-3}$)  & for $q$ \\
 \\ \hline
\hline
GRS 1915+105    & 33.85  &  0.042(24)  &  {\bf 2$\times10^{-4}$}   &    K1-5  &   2.4-2.9  &    1-4$\times10^{-4}$  & (1)  \\
V404 Cyg             &  6.471 &  0.067(5)    &  {\bf 5$\times10^{-3}$}  &  G8-K4  &    1.7-2.8   &   1-9$\times10^{-4}$   & (2) \\
\hline
BW Cir                  &   2.544 &   0.12(3)   &   {\bf 0.03}  &    G0-5   &  1.1-1.5 &    1-6$\times10^{-3}$  & (3) \\
GX 339-4              &  1.759  &  0.18(5)    &   {\bf 0.06}  &    K1-2    &   2.4  &      3-4$\times10^{-4}$  & (4) \\
XTE J1550-564    &  1.542  &  0.033(8)   &   {\bf 0.08}  &  K2-4    &  2.4-2.8   &  1-3$\times10^{-4}$  & (5) \\
\hline
MAXI J1820+070 &   0.685  &   0.072(12) &  {\bf 0.41}  &   K3-5  &  2.6-2.9  &   1-2$\times10^{-4}$  & (6) \\
N Oph 77             &   0.523  &   0.014(16)  &  {\bf 0.65} &   K1-5  &  2.4-2.9   &  1-4$\times10^{-4}$  & (7) \\
N Mus 91             &   0.433  &    0.079(7)  &  {\bf 1.03}  &   K0-6  &  2.3-3.3   &  1-4$\times10^{-4}$  & (8) \\
MAXI J1305-704  &   0.394  &   0.05(2)     &  {\bf 1.22}   &   K3-5   &  2.6-2.9   &  1-2$\times10^{-4}$  & (9) \\
GS 2000+25         &   0.344  &   0.042(12)   &   {\bf 1.58}   &  K3-7    &   2.6-3.6   &  0.7-2$\times10^{-4}$ & (10)  \\
A 0620-00             &   0.323  &    0.067(10)  &   {\bf 1.83}  &   K0-7    &  2.3-3.6    &   0.7-4$\times10^{-4}$  & (11)  \\
XTE J1650-500    &   0.321  &     0.03(3)     &   {\bf 1.80}   &   K1-7  &   2.4-3.6  &      0.7-4$\times10^{-4}$  & (12)  \\
N Vel 93               &   0.285   &   0.055(10)  &   {\bf 2.33}   &  K0-M0   &   2.3-4.0  &    0.3-4$\times10^{-4}$  & (13)  \\
XTE J1859+226   &   0.276    &  0.07(1)      &   {\bf 2.52}    &  K5-7    &  2.9-3.6    &     0.7-1$\times10^{-4}$  & (14)  \\
GRO J0422+32    &   0.212    &  0.12(8)      &   {\bf 4.35}    &  M0-5   &   4.0-30    &   $\lesssim$3$\times10^{-5}$  & (15)  \\
XTE J1118+480    &   0.170    & 0.024(9)    & {\bf 6.32}   &  K5-M1 &    2.9-5.6   &    0.3-1$\times10^{-4}$  & (16)  \\
SWIFT J1753.5-0127  &  0.136   & 0.027(3)   & {\bf 9.96}   &  M4-5 &    16-30   &   $\lesssim$3$\times10^{-5}$   & (17)  \\
\hline
\end{tabular}
\tablefoot{\\
\tablefoottext{a}{Widest spectral range given by Table~\ref{table:b1}.}
\tablefoottext{b}{MS densities derived from the compilation of masses and radii of \citet{pecaut13}.}
\tablefoottext{c}{Giant densities from \citet{drilling02}.}
}

\tablebib{(1)~\citet{steeghs13}; (2) \citet{casares96}; (3) \citet{casares09}; (4) \citet{heida17};
(5) \citet{orosz11}; (6) \citet{torres20}; (7) \citet{harlaftis97}; (8) \citet{wu15}; (9) \citet{mata21}; 
(10) \citet{harlaftis96}; (11) \citet{marsh94}; (12) \citet{casares96}; (13) \citet{macias11};  (14) \citet{yanes22}; 
(15) \citet{harlaftis99};  (16) \citet{gonzalez14}; (17) \citet{yanes25}. 
}
\end{table*}

 \FloatBarrier

Figure~\ref{ap:porb_teff} shows that LMXB donors with $P_{\rm orb}\gtrsim$1.5 d have crossed the TAMS line and follow the 
stripped-giant branch, while those with $P_{\rm orb}\lesssim$1.5 d have not left the MS.  In particular, donors with 
$P_{\rm orb}\lesssim$  0.3 d (=7 h) are only slightly evolved  compared to MS stars on the empirical track of \citet{pecaut13}.   
XTE J1118+480 is a clear outlier since it is $\sim$500 K  hotter than expected for a MS at its orbital period. This is reminiscent 
of the cataclysmic variables QZ Ser, EI Psc and SDSS J1702+3229, whose donors are thought to descend from more massive 
progenitors that were significantly evolved at the onset of mass transfer \citep{thorstensen02a,thorstensen02b,littlefair06}. 
Interestingly, XTE J1118+480 has shown evidence for CNO processed material, in support for a highly evolved donor progenitor 
\citep{haswell02}. 

Overall, LMXB donors are seen to follow a well-defined path in the $P_{\rm orb}-T_{\rm eff}$  plane. Unfortunately, this is weakly 
constrained at short orbital periods $\lesssim$0.14 d due to lack of data. To compensate for this we decided to include an updated 
collection of CV donor spectral types from \cite{knigge06}. These have been converted to $T_{\rm eff}$ values, as was previously 
done for BH LMXBs. The addition of CV information seems justified, since previous work has shown that, for the same orbital period, 
donor stars in CVs and LMXBs  are almost indistinguishable (e.g. \citealt{smith98}). 
 
In order to characterise the $T_{\rm eff}$ evolution with $P_{\rm orb}$ we fit the ensamble of BH LMXB and CV data points in three 
different regions: (i) short-periods below the period gap $P_{\rm orb}\leq0.085$ d (ii) intermediate-periods  0.085 d $\leq P_{\rm orb}\leq0.2$ d  
and (iii) long-periods $P_{\rm orb}\geq0.2$ d. We find that a broken power-law provides a good description of the long-period systems, 
while simple linear fits are used for shorter periods i.e. 

\begin{equation}T_{\rm eff} = \left\{
\begin{array}{l l}
 1810+14611 \times P_{\mathrm{orb}} & P_{\rm orb} < 0.085 ~{\rm d} \\
 2856+2732 \times P_{\mathrm{orb}} & 0.085~{\rm d}\leq P_{\rm orb} < 0.213~{\rm d} \\
 4645-1.09 \times P_{\mathrm{orb}}^{~-4.52} & 0.213~{\rm d} \leq P_{\rm orb}   \\
 \end{array}
  \right.
\label{eq:teff_porb}
\end{equation}

\noindent
 These empirical fits are represented by the red line in  Fig.~\ref{ap:porb_teff}.  For the reasons given above, the five outliers 
 (BW Cir, XTE J1118+480 plus the three CVs with confirmed nuclear-evolved donors) were masked in  the fits.  We note, 
 however, that their inclusion does not significantly alter the results of the fits. 

The surprisingly narrow path traced by LMXB donors in the $P_{\rm orb}-T_{\rm eff}$ plane suggests that they all follow 
the same evolutionary track. \citet{king96} showed that systems with $P_{\rm orb}\lesssim$2 d evolve towards shorter periods 
because angular momentum losses shrink the binary orbit faster than stellar expansion. Conversely, for $P_{\rm orb}\gtrsim$2 d  
the companion is nuclear-evolved before the onset of mass transfer and the binary evolves to increasing orbital periods 
(see also \citealt{pylyser88}). The gap seen at $\simeq 0.7-1.5$ d thus reflects a real shortage of systems triggered by the 
bifurcation period, which causes binaries to evolve towards either shorter or longer $P_{\rm orb}$. In addition,  the small 
scatter seen at periods $\approx$0.25-1 d implies that the orbital separation after common envelope ejection must have been 
sufficiently tight for the donor stars to come into contact before evolving significantly. To obtain a better description of the 
$P_{\rm orb}-T_{\rm eff}$  relation for BH LMXBs at very short orbital periods $\lesssim$0.15 d, it would be important to 
determine spectral types for the donor stars in Swfit J1357.2-0933 (0.12 d) and MAXI J1659-152 (0.10 d). This will require 
infrared spectroscopy since these stars are too cold to be detected at visible wavelengths \citep{mata15,torres15,torres21}.

\section{ A simulation of \ha~EWs in quiescent BH X-ray transients}
\label{ap:model_ew}

We have built a toy model to simulate the EW of the \ha~emission in quiescent BH XRTs. 
We assume typical BH XRT parameters, with a compact object mass $M_{1}=8$ \msun~and mass ratio 
$q=0.06$. The binary separation is set by Kepler's Third law $a\propto (P_{\rm orb}^{2} M_{1} (1+q) )^{1/3}$ 
and the size of the secondary star by the effective Roche lobe radius through Eggleton's approximation 
$R_{2}/a=0.49~q^{2/3} /\left[(0.6~q^{2/3} + \ln (1+q^{2/3})\right]$ \citep{eggleton83}. The accretion disc is 
simulated by a flat cylinder truncated at the 3:1  resonance radius $r_{d}/a=(1/3)^{2/3} (1+q)^{-1/3} $ \citep{frank02}.  
Based on observations of extreme wing velocities in \ha~profiles we set  the inner disc radius at $r_{in}=0.06 r_{d}$ 
\citep{casares22}, although this parameter has very little impact on the results of the simulation.  

To estimate the EW of the \ha~line we start by computing the continuum emitted by the secondary star and 
the accretion disc using blackbody approximations. The flux density radiated by the secondary star, $f_s (\lambda)$,
 is obtained from a blackbody with radius $R_2$ and temperature $T_{\rm eff}$, where $T_{\rm eff} $ depends on 
 $P_{\rm orb}$ according to eq.~\ref{eq:teff_porb}. For the flux density of the accretion disc continuum, $f_d (\lambda)$, 
 we adopt a multi-colour blackbody with a flat temperature profile $T_{r}=T_{in} (r/ r_{in})^{-0.25}$ and inner disc temperature 
 $T_{in}=4500$ K, which are appropriate choices for the quiescent state \citep{orosz97,beer02}. In order to reproduce the 
 far ultraviolet excess widely observed in quiescent BH XRTs \citep{mcclintock95, mcclintock03,froning11,hynes12, poutanen22} 
we also include emission from a hot blackbody, $f_h (\lambda)$, with temperature $T_{h}=10000$ K and size $r_{h}=r_{in}$. 
This component has been attributed to the transition region between the thin disc and an advection dominated flow, although 
an origin on the bright spot, where the gas stream hits the outer disc, cannot be excluded (see \citealt{froning11, hynes12}). 

\begin{figure}
  \centering
    \includegraphics[angle=-90,width=\columnwidth]{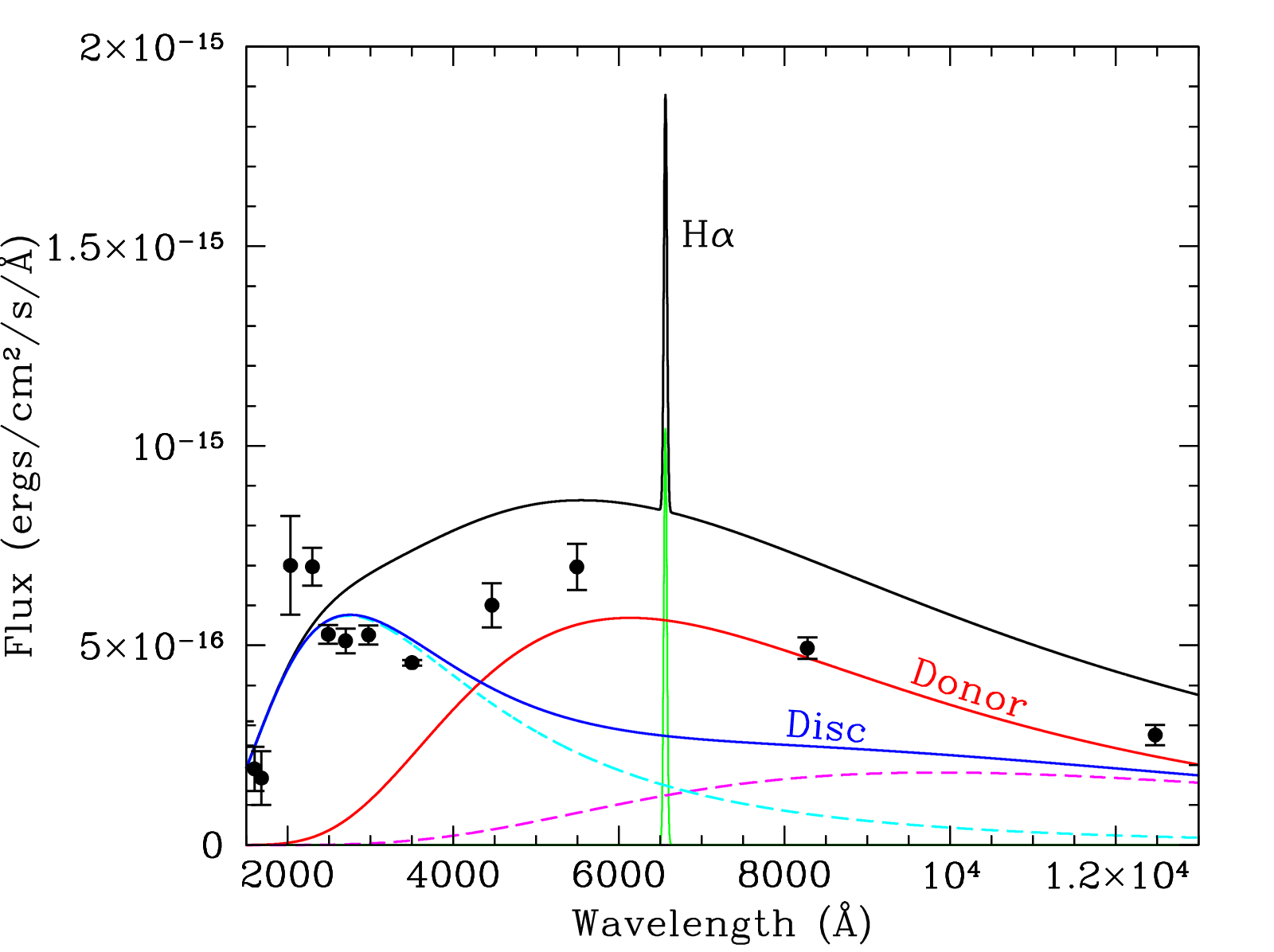}
      \caption{SED of A0620-00 simulated with our toy model, using the physical parameters of \cite{cantrell10}. 
      The accretion disc contribution to the SED (blue)  consists of a multicolour blackbody (dashed magenta) 
      plus an inner hot-spot with $T_{h}=10000$ K (dashed cyan). The flux of the \ha~line (green Gaussian) 
      has been scaled so that EW=58 \AA.  
      For comparison, we overlay SED photometric points from \cite{froning11}. 
                       }
         \label{ap:sed_a0620}
   \end{figure}

To simulate the \ha~flux we assume optically thin emission from the surface of the accretion disc. We approximate the 
\ha~profile by a Gaussian function with an integrated flux $F_{H\alpha}$  which is proportional to the area of the accretion 
disc $\pi (r_{d}^{2}-r_{in}^{2})$. $F_{H\alpha}$ has been scaled to give EW=58 \AA~for the binary parameters of the 
canonical BH XRT A0620-00. The EW of the \ha~line  is then calculated as 

\begin{equation}
EW(H\alpha)={\frac{F_{H\alpha} }{ f_s(\lambda_0)  + \cos i  \left[ f_d(\lambda_0) + f_h(\lambda_0) \right] }} 
\label{eq:ew}
\end{equation}

\noindent
with $\lambda_{0}=6563$ \AA. The $\cos i$ factor accounts for the foreshortening of the accretion disc flux caused by 
binary inclination. In this crude approximation we neglect limb darkening as it has little effect on our limited  
$\lambda$ and $T_{\rm eff}$ range of interest. As an example, Fig.~\ref{ap:sed_a0620} presents the simulated spectral 
energy distribution (SED) for the case of A0620-00, where we adopt $P_{\rm orb}=0.323$ d, $q=0.06$, $M_{1}=7$ \msun, 
$i=53^{\circ}$ and $d=1.1$ kpc \citep{cantrell10}. 
For reference, we also overlay NUV, optical and NIR photometric datapoints from \cite{froning11}, 
dereddened with $E(B-V)=0.35$ and $R_{V}=3.1$.  Our toy model provides a reasonable representation 
of the observed SED, taking into account the overarching simplifications, uncertainties in system parameters  (e.g. the Gaia DR2 parallax gives $d=1.6\pm0.4$;  
kpc \citealt{gandhi19}), and intrinsic variability common to quiescent XRTs \citep{cantrell08}. In any case, we emphasise that the simulation is not intended to be 
an accurate representation of the full SED for a given system, but an attempt to explore the dependence of the \ha~EW on binary 
parameters in a statistical way. 
 
\begin{figure} 
  \centering
    \includegraphics[angle=0,width=\columnwidth]{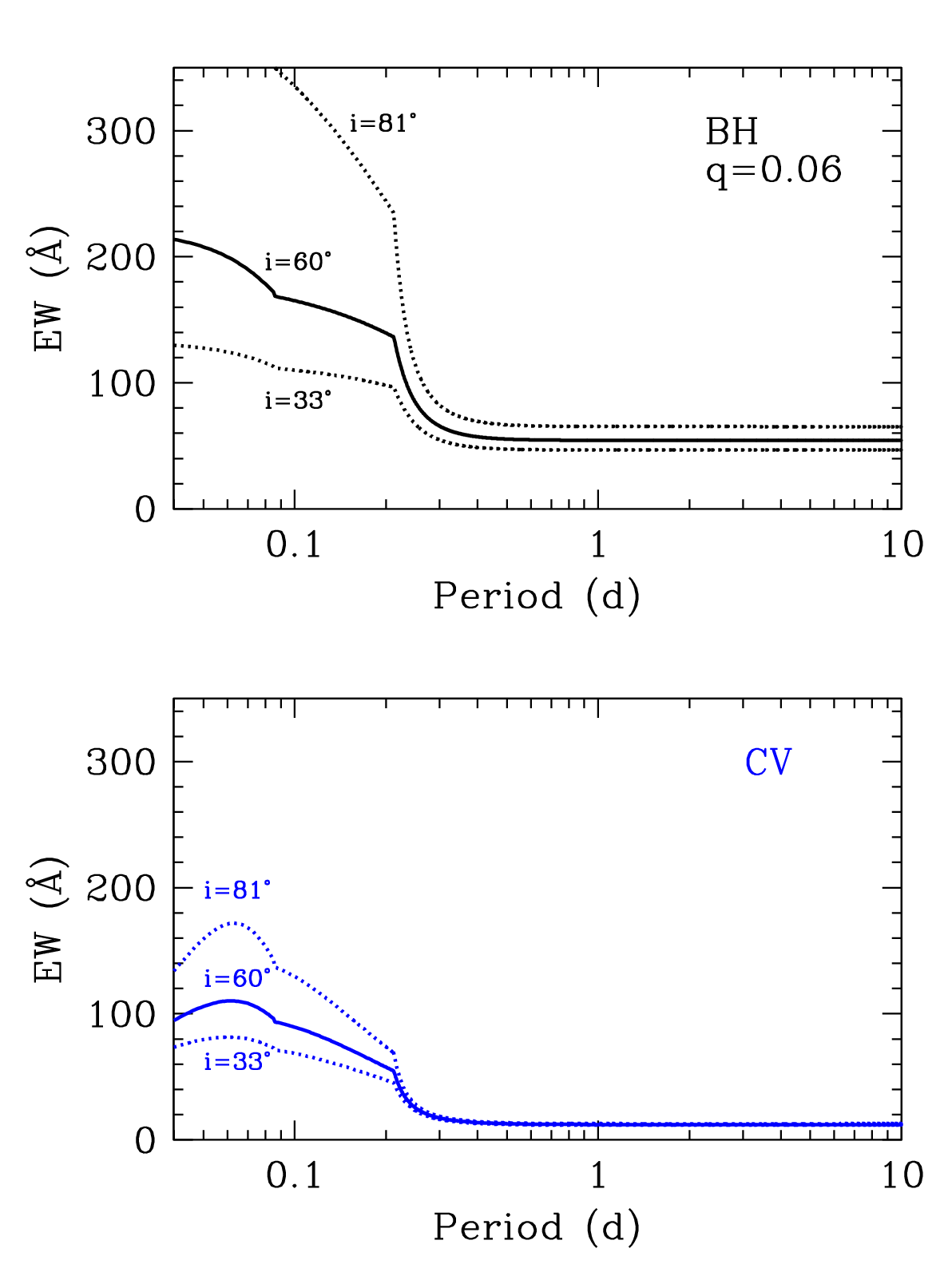}
      \caption{Simulated EWs as a function of orbital period for BH XRTs (top panel) and CVs (bottom panel). In the latter 
      case we add the contribution of a white dwarf with $T_{\rm WD}=15000$ K and $r_{\rm WD}=0.01$ \rsun~to the 
      continuum. Also, the $P_{\rm orb}-q$ dependence, derived in C18, is applied. Three different inclinations are represented. 
        }
         \label{ap:porb_ew}
   \end{figure}

Since $M_1$  and the distance to the object cancel out in eq.~\ref{eq:ew} our synthetic EWs depend only on 
$P_{\rm orb}$, $q$ and inclination. This is illustrated in the top panel of Fig.~\ref{ap:porb_ew}, which represents 
the evolution of the EW with $P_{\rm orb}$ for a typical BH XRT with $q=0.06$ and three different inclinations: 
$i=33^{\circ}$, 60$^{\circ}$ and 81$^{\circ}$ i.e. the median and $\pm 1 \sigma$ of the isotropic distribution. 
The figure shows that the EW is mostly determined by changes in the companion's temperature with  $P_{\rm orb}$ 
(see Fig.~\ref{ap:porb_teff}). Only at short periods $P_{\rm orb}\lesssim0.2$ d the drop in companion temperature 
does cause the disc contribution to start dominating the continuum flux. This amplifies the effect of inclination in 
reducing the disc brightness and, therefore, enlarges the range of possible EW values.

For comparison, we also simulate the evolution of the EW for the case of CVs (bottom panel in Fig.~\ref{ap:porb_ew}). 
CVs have less extreme mass ratios, with typical values ranging between $q\simeq0.1-1$ and a mean at $q\simeq0.6$ 
\citep{ritter03}. Since mass ratio is known to increase with $P_{\rm orb}$ we apply here the  relation derived in C18, 
i.e. $q=0.73-11.55\times \exp{\left[-(P_{\rm orb}+0.39)/0.15\right]}$. We also include an additional blackbody to account 
for the white dwarf emission, with fiducial parameters $T_{\rm WD}=15000$ K and $r_{\rm WD}=0.01$ \rsun~\citep{savoury11}.
The figure shows that EWs are systematically lower in CVs than in BH XRTs, a natural consequence of their larger $q$ values, 
which increase the relative contribution of the companion to the total flux. Only below the period gap $P_{\rm orb}\leq0.085$ d, 
CV mass ratios begin to compare to those in BH XRTs, but then the contribution of the white dwarf continuum becomes important, 
capping the observed EWs. In summary, our simulation predicts that the combined effect of large mass ratios and dilution by the 
white dwarf continuum (most important at very short $P_{\rm orb}$), leads to smaller EWs for CVs than for BH XRTs.

\end{appendix}


\begin{thebibliography}{}

\bibitem[Abbott et al.(2019)]{abbott19}
Abbott B.P., Abbott,  R., Abbott, T.D., et al. 2019, \apj, 882, L24

\bibitem[Arur \& Maccarone(2018)]{arur18}
Arur, K., \& Maccarone, T.J. 2018, \mnras, 474, 69

\bibitem[Baglio et al. (2023)]{baglio23}
Baglio, M.C., Russell, D.M., Alabarta, K., et al. 2023, ATel., 16192, 1 

\bibitem[Beer \& Podsiadlowski (2002)]{beer02}
Beer, M.E., \& Podsiadlowski, P. 2002, \mnras, 331, 351 

\bibitem[Bellm et al. (2023)]{bellm23}
Bellm, E.C., Wang, Y., van Roestel, J., et al. 2023, \apj, 956, 21  

\bibitem[Bernardini et al. (2016)]{bernardini16}
Bernardini, F., Russell, D.M., Shaw, A.W., et al. 2016, \apj, 818, L5

\bibitem[Bertelli et al. (2008)]{bertelli08}
Bertelli, G., Girardi, L., Marigo, P., \& Nasi, E. 2008, \aap, 484, 815

\bibitem[Bertelli et al. (2009)]{bertelli09}
Bertelli, G., Nasi, E., Girardi, L., \& Marigo, P. 2009, \aap, 508, 355 

\bibitem[Burdge et al. (2024)]{burdge24}
Burdge, K.B., El-Badry, K., Kara, E., et al. 2024, \nat, 635, 316 

\bibitem[Buxton \& Vennes (2001)]{buxton01}
Buxton, M., \& Vennes, S. 2001, \pasa, 18, 91 

\bibitem[Callanan \& Charles (1991)]{callanan91}
Callanan, P.J., \& Charles, P.A. 1991, \mnras, 249, 573

\bibitem[Calvelo et al.(2009)]{calvelo09}
Calvelo, D.E., Vrtilek, S.D., Steeghs, D., et al. 2009, \mnras, 399, 539  

\bibitem[Cantrell et al.(2008)]{cantrell08}
Cantrell, A.G., Bailyn, C.D., McClintock, J.E., \& Orosz, J.A., 2008, \apj, 673, L159  

\bibitem[Cantrell et al.(2010)]{cantrell10}
Cantrell, A.G., Bailyn, C.D., Orosz, J.A., et al., 2010, \apj, 710, 1127  

\bibitem[Casares (1996)]{casares96} 
Casares, J. 1996, Astrophysics and Space Science Library, Proc. of the  
158th coll. of IAU, ed. A. Evans \& J.H. Wood, Dordrecht: Kluwer Academic Publishers,  
Vol. 208, p.395

\bibitem[Casares (2015)]{casares15}
Casares, J. 2015, \apj, 808, 80)

\bibitem[Casares (2016)]{casares16}
Casares, J. 2016, \apj, 822, 99

\bibitem[Casares (2018)]{casares18}
Casares, J. 2018, \mnras, 473, 5195

\bibitem[Casares \& Charles (1994)]{casares94}
Casares, J., \& Charles, P.A. 1994, \mnras, 271, L5 

\bibitem[Casares \& Torres (2018)]{casares-torres18}
Casares, J., \& Torres, M.A.P. 2018, \mnras, 481, 4372 

\bibitem[Casares et al.(1993)]{casares93}
Casares, J., Charles, P.A., Naylor, T., \& Pavlenko, E.P. 1993, MNRAS, 265, 834  

\bibitem[Casares et al.(1995a)]{casares95a}
Casares, J., Charles, P.A., \& Marsh, T.R., 1995a, \mnras, 277, L45  

\bibitem[Casares et al.(1995b)]{casares95b}
Casares, J., Martin, A.C., Charles, P.A., et al. 1995b, \mnras, 276, L35  

\bibitem[Casares et al.(2009)]{casares09}
Casares, J., Orosz, J.A., Zurita, C., et al. 2009, \apjs, 181, 238

\bibitem[Casares et al.(1997)]{casares97}
Casares, J., Mart\'\i{}n, E.L., Charles, P.A., Molaro, P., \& Rebolo, R. 1997, \na, 1, 299

\bibitem[Casares et al. (2004)]{casares04}
Casares, J., Zurita, C., Shahbaz, T., Charles, P.A., \& Fender, R.P. 2004, \apj, 613, L133

\bibitem[Casares et al. (2009)]{casares09}
Casares, J., Orosz, J.A., Zurita, C., et al. 2009, \apjs, 181, 238

\bibitem[Casares et al. (2019)]{casares19}
Casares, J.,, Mu\~noz-Darias, T.,  Mata S\'anchez, D., et al.  2019, \mnras, 488, 1356

\bibitem[Casares et al. (2022)]{casares22}
Casares, J.,, Torres, M.A.P., Mu\~noz-Darias, et al.  2022, \mnras, 516, 2023

\bibitem[Casares et al. (2023)]{casares23}
Casares, J.,, Yanes-Rizo, I.V., Torres, M.A.P., et al.  2023, \mnras, 526, 5209

\bibitem[Chevalier et al. (1999)]{chevalier99}
Chevalier, C., Ilovaisky, S.A., Leisy, P., \& Patat, F. 1999, \aap, 347, L51

\bibitem[Ciatti \& Vittone (1977)]{ciatti77}
Ciatti, F., \& Vittone, A. 1977, IBVS N. 1261
 
\bibitem[Corral-Santana et al. (2011)]{corral11}
Corral-Santana, J.M., Casares, J., Shahbaz, T., et al. 2011, \mnras, 413, L15

\bibitem[Corral-Santana et al.(2013)]{corral13}
Corral-Santana, J.M., Casares, J., Mu\~noz-Darias, T., et al., 2013, Science, 339, 1048

\bibitem[Corral-Santana et al. (2016)]{corral16}
Corral-Santana, J.M., Casares, J.,  Mu\~noz-Darias, T.,  Bauer, F.E., Mart\'\i{}nez-Pais, I.G. \& Russell, D.M 2016, \aap, 587, A61

\bibitem[Corral-Santana et al. (2018)]{corral18}
Corral-Santana, J.M., Torres, M.A.P., Shabaz, T., et al. 2018, \mnras, 475, 1036 

\bibitem[Drilling \& Landolt(2002)]{drilling02}
Drilling, J.S., \& Landolt, A.U. 2002, {\it Normal Stars} in Allen's Astrohysical Quantities, ed. A.N. Cox, Springer New York , NY,  p. 381,  
ISBN 978-0-387-95189-8

\bibitem[Eggleton(1983)]{eggleton83}
Eggleton, P.P. 1983, \apj, 268, 368

\bibitem[Filippenko et al. (1995)]{filippenko95}
Filippenko, A.V., Matheson, T., \& Barth, A.J. 1995, \apj, 455, L139

\bibitem[Filippenko et al. (1997)]{filippenko97}
Filippenko, A.V., Leonard, D.C., Matheson, T., et al. 1999, \pasp, 111, 969

\bibitem[Filippenko et al. (1999)]{filippenko99}
Filippenko, A.V., Leonard, D.C., Matheson, T., et al. 1999, \pasp, 111, 969

\bibitem[Frank et al.(2002)]{frank02}
Frank, J., King, A.R., \& Raine, D.J. 2002, Accretion Power in Astrophysics, Vol.
21 (3rd ed.; Cambridge: Cambridge Univ. Press)

\bibitem[Froning et al. (2007)]{froning07}
Froning, C.S., Robinson, E.L., \& Bitner, M.A. 2007, \apj, 663, 1215 
     
\bibitem[Froning et al. (2011)]{froning11}
Froning, C.S., Cantrell, A.G., Maccarone, T.J., et al. 2011, \apj, 743, 26 

\bibitem[Gaia Collaboration et al.(2024)]{gaia24}
Gaia Collaboration, et al. 2024, \aap, 686, L2

\bibitem[Gandhi et al.(2019)]{gandhi19}
Gandhi, P., Rao, A., Johnson,  M.A.C., Paice, J.A., \& Maccarone, T.J. 2019, \mnras, 485, 2642 

\bibitem[Gandhi et al.(2020)]{gandhi20}
Gandhi, P., Rao, A., Charles,  P.A., et al. 2020, \mnras, 496, L22 

\bibitem[Garcia \& Wilkes(2002)]{garcia02}
Garcia, M.R., \& Wilkes, B.J. 2002, ATel., 104, 1

\bibitem[Garcia et al.(1996)]{garcia96}
Garcia, M.R., Callanan, P.J., McClintock, J.E., \& Zhao, P. 1996, \apj, 460, 932 

\bibitem[G\"ansicke et al.(2009)]{gansicke09}
G\"ansicke, B.T., Dillon, M., Southworth, J., et al. 2009, \mnras, 397, 2170

\bibitem[Gelino et al. (2001)]{gelino01}
Gelino, D.M., Harrison, T.E., \& Orosz, J.A. 2001, \aj, 122, 2668 
     
\bibitem[Gelino \& Harrison (2003)]{gelino03}
Gelino, D.M., \& Harrison, T.E. 2003, \apj, 599, 1254 

\bibitem[Gelino et al. (2006)]{gelino06}
Gelino, D.M., Balman, \c{S}., Kizilo\u{g}lu, \"U., et al. 2006, \apj, 642, 438

\bibitem[Gonz\'alez Hern\'andez et al.(2004)]{gonzalez04}
Gonz\'alez Hern\'andez J.I., Rebolo, R., Israelian, G., et al. 2004, \apj, 609, 988

\bibitem[Gonz\'alez Hern\'andez et al.(2006)]{gonzalez06}
Gonz\'alez Hern\'andez J.I., Rebolo, R., Israelian, G., et al. 2006, \apj, 644, L49

\bibitem[Gonz\'alez Hern\'andez et al.(2008a)]{gonzalez08a}
Gonz\'alez Hern\'andez J.I., Rebolo, R., \& Israelian, G. 2008a, \aap, 478, 203

\bibitem[Gonz\'alez Hern\'andez et al.(2008b)]{gonzalez08b}
Gonz\'alez Hern\'andez, J.I., Rebolo, R., Israelian, G., et al.  2008b, \apj, 679, 732

\bibitem[Gonz\'alez Hern\'andez \& Casares (2010)]{gonzalez10}
Gonz\'alez Hern\'andez J.I., \& Casares J., 2010, \aap, 516, A58

\bibitem[Gonz\'alez Hern\'andez et al.(2011)]{gonzalez11}
Gonz\'alez Hern\'andez J.I., Casares J., Rebolo, R., et al. 2011, \apj, 738, 95

\bibitem[Gonz\'alez Hern\'andez et al.(2012)]{gonzalez12}
Gonz\'alez Hern\'andez J.I., Rebolo, R., \& Casares, J.  2012, \apj, 744, L25

\bibitem[Gonz\'alez Hern\'andez et al.(2014)]{gonzalez14}
Gonz\'alez Hern\'andez J.I., Rebolo, R., \& Casares, J.  2014, \mnras, 438, L21

\bibitem[Gonz\'alez Hern\'andez et al.(2017)]{gonzalez17}
Gonz\'alez Hern\'andez J.I., Su\'arez-Andr\'es, L., Rebolo, R., \& Casares, J.  2017, \mnras, 465, L15

\bibitem[Harlaftis et al. (1996)]{harlaftis96}
Harlaftis, E.T., Horne, K., \& Filippenko, A.V. 1996 \pasp, 108, 762

\bibitem[Harlaftis et al. (1997)]{harlaftis97}
Harlaftis, E.T., Steeghs, D., Horne, K., \& Filippenko, A.V. 1997 \aj, 114, 1170

\bibitem[Harlaftis et al. (1999)]{harlaftis99}
Harlaftis, E.T., Collier, S., Horne, K., \& Filippenko, A.V. 1999 \aap, 341, 491

\bibitem[Harlaftis \& Greiner (2004)]{harlaftis04}
Harlaftis, E.T., \& Greiner, J. 2004, \aap, 414, L13

\bibitem[Harrison et al. (2007)]{harrison07}
Harrison, T.E., Howell, S.B., Szkody, P., \& Cordova, F.A. 2007, \aj, 133, 162

\bibitem[Haswell et al. (2002)]{haswell02}
Haswell, C.A., Hynes, R.I., King, A.R., \& Schenker, K. 2002, \mnras, 332, 928

\bibitem[Heida et al. (2017)]{heida17}
Heida, M., Jonker, P.G., Torres, M.A.P., \& Chiavassa, A. 2017, \apj, 846, 132

\bibitem[Hynes et al. (2002)]{hynes02}
Hynes, R.I., Zurita, C., Haswell, C.A., et al. 2002, \mnras, 330, 1009 

\bibitem[Hynes et al. (2004)]{hynes04}
Hynes, R.I., Charles, P.A., Garcia, M.R., et al. 2004, \apj, 611, L125

\bibitem[Hynes et al. (2009)]{hynes09}
Hynes, R. I.; Bradley, C. K.; Rupen, M., et al. 2009, \mnras, 399, 2239

\bibitem[Hynes \& Robinson (2012)]{hynes12}
Hynes, R. I. \&  Robinson, E.L. 2012, \apj, 749, 3

\bibitem[Inight et al.(2023)]{inight23}
Inight, K., G\"ansicke, B.T., Breedt, E., et al. 2023, \mnras, 524, 4867

\bibitem[Israelian et al. (1999)]{israelian99}
Israelian, G., Rebolo, R., Basri, G., Casares, J., \& Mart\'\i{}n, E.L. 1999, \nat, 401, 142

\bibitem[Jonker et al. (2021)]{jonker21}
Jonker, P.G., Kaur, K., Stone, N., \& Torres, M.A.P. 2021, \apj, 921, 131 

\bibitem[Khargharia et al.(2010)]{khargharia10}
Khargharia, J., Froning, C.S., \& Robinson E.L. 2010, \apj, 716, 1105

\bibitem[Khargharia et al.(2013)]{khargharia13}
Khargharia, J., Froning, C.S., Robinson, E.L., \& Gelino, D.M. 2013, \aj, 145, 21

\bibitem[King (1993)]{king93}
King, A.R. 1993, \mnras, 260, L5

\bibitem[King, Kolb \& Burderi (1996)]{king96}
King, A.R., Kolb, U., \& Burderi, L. 1996, \apj, 464, L127

\bibitem[King et al. (1996)]{king96b}
King, N.L., Harrison, T.E., \& McNamara, B.J. 1996, \aj, 111, 1675

\bibitem[Knevitt et al.(2014)]{knevitt14}
Knevitt, G., Wynn, G.A., Vaughan, S. \& Watson, M.G. 2014, \mnras, 437, 3087

\bibitem[Knigge (2006)]{knigge06}
Knigge, C. 2006, \mnras, 373, 484

\bibitem[Koljonen et al. (2016)]{koljonen16}
Koljonen, K.I.I., Russell, D.M., Corral-Santana, J.M., et al. 2016, \mnras, 460, 942

\bibitem[Kreidberg et al.(2012)]{kreidberg12}
Kreidberg, L., Bailyn, C.D., Farr, W. ,\& Kalogera, V. 2012, \apj, 757, 36 

\bibitem[Kuulkers et al. (2013)]{kuulkers13}
Kuulkers, E., Kouveliotou C., Belloni, T., et al. 2013, \aap, 552, A32

\bibitem[Lin et al.(2019)]{lin19}
Lin, J., Yan, Z., Han, Z.,\& Yu, W. 2019, \apj, 870, 126

\bibitem[Littlefair et al.(2006)]{littlefair06}
Littlefair, S., Dhillon, V.S., Marsh, T.R., \& G\"ansicke, B.T. 2006, \mnras, 371, 1435

\bibitem[Maccarone \& Patruno(2013)]{maccarone13}
Maccarone, T.J., \& Patruno, A., 2013 \mnras, 428, 1335

\bibitem[Macias et al.(2011)]{macias11}  
Macias, P., Orosz, J.A., Bailyn, C.D.,  et al. 2011, \mbox{Bulletin of the American Astronomical Society}, Vol. 43, 2011

\bibitem[Marsh et al.(1994)]{marsh94}
Marsh, T.R.,  Robinson, E.L., \& Wood, J.H. 1994, \mnras, 266, 137

\bibitem[Mata S\'anchez et al.(2015)]{mata15}
Mata S\'anchez, D., Mu\~noz-Darias, T., Casares, J., Corral-Santana, J.M., \& Shahbaz, T. 2015, \mnras, 454, 2199

\bibitem[Mata S\'anchez et al.(2017)]{mata17}
Mata S\'anchez, D., Mu\~noz-Darias, T., Casares, J., \& Jim\'enez Ibarra, F. 2017, \mnras, 464, L41

\bibitem[Mata S\'anchez et al.(2021)]{mata21}
Mata S\'anchez, D., Rau, A., \'Alvarez-Hern\'andez, A., et al. 2021, \mnras, 506, 581

\bibitem[Menou et al.(1999)]{menou99}
Menou, K., Narayan, R., \& Lasota, J.-P. 1999, \apj, 513, 811

\bibitem[McClintock \& Remillard (1986)]{mcclintock86}
McClintock, J.E., \& Remillard, R.A. 1986, \apj, 308, 110

\bibitem[McClintock et al. (1995)]{mcclintock95}
McClintock, J.E., Horne, K., \& Remillard, R.A. 1995, \apj, 442, 358

\bibitem[McClintock et al. (2001)]{mcclintock01}
McClintock, J.E., Garcia, M.R., Caldwell, N., et al. 2001, \apj, 551, L147

\bibitem[McClintock et al. (2003)]{mcclintock03}
McClintock, J.E., Narayan, R., Garcia, M.R., et al. 2003, \apj, 593, 435

\bibitem[Murdin et al.(1977)]{murdin77}
Murdin, P., Griffiths, R.E., Pounds, K.A., Watson, M.G., \& Longmore, A.J. 1977, \mnras, 178, 27

\bibitem[Murdin et al.(1980)]{murdin80}
Murdin, P., Allen, D.A., Morton, D.C., Whelan, J.A.J., \& Thomas, R.M. 1980, \mnras, 192 709 

\bibitem[Naoz et al. (2016)]{naoz16}
Naoz, S., Fragos, T., Geller, A., et al. 2016, \apj, 822, L24

\bibitem[Narayan \& McClintock (2005)]{narayan05}
Narayan, R.,  \& McClintock, J.E. 2005, \apj, 623, 1017

\bibitem[Neilsen et al. (2008)]{neilsen08}
Neilsen, J., Steeghs, D., \& Vrtilek, S.D. 2008, \mnras, 384, 849

\bibitem[Oke (1977)]{oke77}
Oke, J.B. 1977, \apj, 217, 181 

\bibitem[Orosz (2003)]{orosz03}
Orosz, J.A. 2003, in A Massive Star Odyssey, from Main Sequence to Supernova, Proc. of IAU Symp., 
ed. K. van der Hucht, A. Herrero, \& C. Esteban (San Francisco: Astronomical Society of the Pacific) Vol 212, p. 3650

\bibitem[Orosz \& Bailyn (1995)]{orosz95}
Orosz, J.A., \& Bailyn, C.D. 1995, \apj, 446, L59 

\bibitem[Orosz \& Bailyn (1997)]{orosz97}
Orosz, J.A., \& Bailyn, C.D. 1997, \apj, 477, 876 
     
\bibitem[Orosz et al.(1994)]{orosz94}
Orosz, J.A., Bailyn, C.D., Remillard, R.A. , McClintock, J.E., \& Foltz, CB.. 1994, \apj, 436, 848

\bibitem[Orosz et al.(1996)]{orosz96}
Orosz, J.A., Bailyn, C.D., McClintock, J.E., \& Remillard, R.A. 1996, \apj, 468, 380

\bibitem[Orosz et al.(1998)]{orosz98}
Orosz, J.A., Jain, R.K., Bailyn, C.D., McClintock, J.E., \& Remillard, R.A. 1998, \apj, 499, 375

\bibitem[Orosz et al.(2001)]{orosz01}
Orosz, J.A., Kuulkers, E., van der Klis, M., et al. 2001, \apj, 555, 489

\bibitem[Orosz et al.(2002)]{orosz02}
Orosz, J.A., Groot, P.J., van der Klis, M., et al. 2002, \apj, 568, 845

\bibitem[Orosz et al.(2004)]{orosz04}
Orosz, J.A., McClintock, J.E., Remillard, R.A., \& Corbel, S. 2004, \apj, 616, 376

\bibitem[Orosz et al.(2011)]{orosz11}
Orosz, J.A., Steiner, J.F., McClintock, J.E., et al. 2011, \apj, 730, 75

\bibitem[{\"O}zel et al.(2010)]{ozel10}
{\"O}zel, F., Psaltis, D., Narayan, R., \& McClintock, J.E. 2010, \apj, 725, 1918 

\bibitem[Paczy\'nski (1971)]{paczynski71}
Paczy\'nski B., 1971, \araa, 9, 183

\bibitem[Pala et al.(2020)]{pala20}
Pala, A.F., G\"ansicke, B.T., Breedt, E., et al. 2020, \mnras, 494, 3799

\bibitem[Pecaut \& Mamajek (2013)]{pecaut13}
Pecauti, M.J., \& Mamajek, E.E. 2013, \apjs, 208, 9

\bibitem[Peterson et al. (2004)]{peterson04}
Peterson, B.M., Ferrarese, L., Gilbert, K.M., et al. 2004, \apj, 613, 682

\bibitem[Podsiadlowski et al.(2002)]{podsiadlowski02}
Podsiadlowski, Ph., Rappaport, S., \& Pfahl, E.D.. 2002, \apj, 565, 1107

\bibitem[Podsiadlowski et al.(2003)]{podsiadlowski03}
Podsiadlowski, Ph., Rappaport, S., \& Han, Z. 2003, \mnras, 341, 385

\bibitem[Poutanen et al.(2022)]{poutanen22}
Poutanen, J., Veledina, A., Berdyugin, A.V., et al. 2022, Science, 375, 874

\bibitem[Pylyser \& Savonije (1988)]{pylyser88}
Pylyser, E., \& Savonije, G.J. 1988, \aap, 191, 57

\bibitem[Remillard et al.(1992)]{remillard92}
Remillard, R.A., McClintock, J.E., \& Bailyn, C.D. 1992, \apj, 399, L145

\bibitem[Remillard et al.(1996)]{remillard96}
Remillard, R.A., Orosz, J.A., McClintock, J.E., \& Bailyn, C.D. 1996, \apj, 459, 226

\bibitem[Ritter \& Kolb (2003)]{ritter03}
Ritter, H., \& Kolb, U. 2003, \aap, 404, 301 

\bibitem[Ruiz et al. (2018)]{ruiz18}
Ruiz, M., Shapiro, S.L., \& Tsokaros, A. 2018,  {\it PhRvD}, 97, 021501

\bibitem[Russell et al. (2017)]{russell17}
Russell, D.M., Lewis, F., \& Gandhi, P. 2017, ATel., 10797, 1 

\bibitem[Russell et al. (2018)]{russell18}
Russell, D.M., Qasim, A.A., Bernardini, F., et al. 2018, \apj, 852, 90 

\bibitem[Saikia {et~al.} (2022)]{saikia22}
Saikia, P., Russel, D.M., Baglio, M.C., et al. 2022, \apj, 932, 38 

\bibitem[Savoury {et~al.} (2011)]{savoury11}
Savoury, C.D.J., Littlefair, S.P., Dhillon, V.S., et al. 2011, \mnras, 415, 2025 

\bibitem[S\'anchez-Fern\'andez {et~al.} (2002)]{sanchez02}
S\'anchez-Fern\'andez, C., Zurita, C., Casares, J., et al. 2002, IAUCirc 7989 

\bibitem[Shahbaz {et~al.} (1996)]{shahbaz96}
Shahbaz, T., van der Hooft, F., Charles, P.A., Casares, J., \& van Paradijs, J. 1996, \mnras, 282, L47 

\bibitem[Shahbaz {et~al.} (1999a)]{shahbaz99a}
Shahbaz, T.., Bandyopadhyay, R.M., \& Charles, P.A. 1999a, \aap, 346, 82

\bibitem[Shahbaz {et~al.} (1999b)]{shahbaz99b}
Shahbaz, T.. van der Hooft, F., Casares, J., Charles, P.A., \& van Paradijs, J. 1999b, \mnras, 306, 89

\bibitem[Shahbaz {et~al.} (2013)]{shahbaz13}
Shahbaz, T., Russell, D.M., Zuritas, C., et al. 2013, \mnras, 434, 2696  

\bibitem[Shibata et al. (2019)]{shibata19}
Shibata, M., Zhou, E., Kiuchi, K., et al.  2019,  {\it PhRvD}, 100, 023015

\bibitem[Siegel et al.(2023)]{siegel23}
Siegel, J.C., Kiato, I., Kalogera, V., et al. 2023, \apj, 954, 212

\bibitem[Smith \& Dhillon (1998)]{smith98}
Smith, D.A., \& Dhillon V.S. 1998, \mnras, 301, 767

\bibitem[Steeghs {et~al.} (2013)]{steeghs13}
Steeghs, D., McClintock, J.E., Parsons, S.G., et al. 2013, \apj, 768, 185

\bibitem[Thorstensen {et~al.} (2002a)]{thorstensen02a}
Thorstensen, J.R., Fenton, W.H., Patterson, J.O., et al. 2002a, \apj, 567, L49

\bibitem[Thorstensen {et~al.} (2002b)]{thorstensen02b}
Thorstensen, J.R., Fenton, W.H., Patterson, J.O., et al. 2002b, \pasp 114, 1117

\bibitem[Torres et al.(2004)]{torres04}
Torres, M.A.P., Callanan, P.J., Garcia, M.R., et al. 2004, \apj, 612, 1026

\bibitem[Torres et al.(2015)]{torres15}
Torres, M.A.P., Jonker, P.G., Miller-Jones, J.C.A., et al. 2015, \mnras, 450, 4292

\bibitem[Torres {et~al.} (2019)]{torres19}
Torres, M.A.P., Casares, J., Jim\'enez-Ibarra, F., {et~al.}  2019, \apj, 882, L21

\bibitem[Torres {et~al.} (2020)]{torres20}
Torres, M.A.P., Casares, J., Jim\'enez-Ibarra, F., {et~al.} 2020, \apj, 893, L37

\bibitem[Torres et al.(2021)]{torres21}
Torres, M.A.P., Jonker, P.G., Casares, J., Miller-Jones, J.C.A., \& Steeghs, D. 2021, \mnras, 501, 2174

\bibitem[van Belle et al.(1999)]{vanbelle99}
van Belle, G.T., Lane, B.F., Thompson, R.R., et al. 1999, \aj, 117, 521

\bibitem[Wagner et al. (2001)]{wagner01}
Wagner, R.M., Foltz, C.B., Shahbaz, T., et al. 2001, \apj, 556, 42

\bibitem[Webb et al.(2000)]{webb00}
Webb, N.A., Naylor, T., Ioannou, Z., Charles, P.A., \& Shahbaz, T. 2000, \mnras, 317, 528

\bibitem[Webbink et al.(1983)]{webbink83}
Webbink, R.F.., Rappaport S.A., \& Savonije, G.J. 1983, \apj, 270, 678

\bibitem[Whelan et al.(1977)]{whelan77}
Whelan, J.A.J., Ward, M.J., Allen, D.A., et al. 1977, \mnras, 180, 657 

\bibitem[Wu et al.(2010)]{wu10}
Wu,Y.X., Yu, W., Li, T.P., Maccarone, T.J., \& Li, X.D.  2010, \apj, 718, 620

\bibitem[Wu et al.(2015)]{wu15}
Wu, J., Orosz, J.A., McClintock, J.E., et al. 2015, \apj, 806, 92

\bibitem[Yanes-Rizo et al.(2022)]{yanes22}
Yanes-Rizo, I.V., Torres, M.A.P., Casares, J., et al. 2022, \mnras, 517, 1476

\bibitem[Yanes-Rizo et al.(2024)]{yanes24}
Yanes-Rizo, I.V., Torres, M.A.P., Casares, J., et al. 2024, \mnras, 527, 5949

\bibitem[Yanes-Rizo et al.(2025)]{yanes25}
Yanes-Rizo, I.V., Torres, M.A.P., Casares, J., et al. 2025, \aap, 694, A119  

\bibitem[Zhang et al.(2017)]{zhang17}
Zhang, G., Russell, D.M., Bernardini, F., Gelfand, J.D., \& Lewis, F. 2017, ATel., 10562, 1 

\bibitem[Zheng et al.(2022)]{zheng22}
Zheng, W.-M., Wu, Q., Wu, J., et al. 2022, \apj, 925, 83

\bibitem[Zorotovic et al. (2011)]{zorotovic11}
Zorotovic, M., Schreiber, M.R., \& G\"ansicke, B.T. 2011, \aap, 536, A42 

\bibitem[Zurita et al. (2002a)]{zurita02a}
Zurita, C., Casares J., Shahbaz T., et al. 2002a, \mnras, 333, 791

\bibitem[Zurita et al. (2002b)]{zurita02b}
Zurita, C., S\'anchez-Fern\'andez, C., Casares, J., et al. 2002b, \mnras, 334, 999 

\bibitem[Zurita et al. (2002c)]{zurita02c}
Zurita, C., Casares, J., Mart\'\i{}nez-Pais, I.G., et al. 2002c, IAUC 7868

\bibitem[Zurita et al. (2006)]{zurita06}
Zurita, C., Torres, M.A.P., Steeghs, D., et al. 2006, \apj, 644, 432

\bibitem[Zurita et al. (2015)]{zurita15}
Zurita, C., Corral-Santana, J.M., \& Casares, J. 2015, \mnras, 454, 3351

\bibitem[Zurita et al. (2016)]{zurita16}
Zurita, C., Gonz\'alez Hern\'andez J.I.,  Escorza, A., \& Casares J., 2016, \mnras, 460, 4289

\end{thebibliography}
\end{document}